\documentclass[letterpaper,compsoc,twoside]{IEEEtran}
\usepackage{fixltx2e} 
\usepackage{cmap} 
\usepackage{ifthen}
\usepackage[T1]{fontenc}
\usepackage[utf8]{inputenc}
\usepackage{amsmath}

\usepackage[font={small,it},labelfont=bf]{caption}
\usepackage{float}

\pdfoutput=1
\usepackage{scipy}
\makeatletter
\def\PY@reset{\let\PY@it=\relax \let\PY@bf=\relax%
    \let\PY@ul=\relax \let\PY@tc=\relax%
    \let\PY@bc=\relax \let\PY@ff=\relax}
\def\PY@tok#1{\csname PY@tok@#1\endcsname}
\def\PY@toks#1+{\ifx\relax#1\empty\else%
    \PY@tok{#1}\expandafter\PY@toks\fi}
\def\PY@do#1{\PY@bc{\PY@tc{\PY@ul{%
    \PY@it{\PY@bf{\PY@ff{#1}}}}}}}
\def\PY#1#2{\PY@reset\PY@toks#1+\relax+\PY@do{#2}}

\expandafter\def\csname PY@tok@gd\endcsname{\def\PY@tc##1{\textcolor[rgb]{0.63,0.00,0.00}{##1}}}
\expandafter\def\csname PY@tok@gu\endcsname{\let\PY@bf=\textbf\def\PY@tc##1{\textcolor[rgb]{0.50,0.00,0.50}{##1}}}
\expandafter\def\csname PY@tok@gt\endcsname{\def\PY@tc##1{\textcolor[rgb]{0.00,0.27,0.87}{##1}}}
\expandafter\def\csname PY@tok@gs\endcsname{\let\PY@bf=\textbf}
\expandafter\def\csname PY@tok@gr\endcsname{\def\PY@tc##1{\textcolor[rgb]{1.00,0.00,0.00}{##1}}}
\expandafter\def\csname PY@tok@cm\endcsname{\let\PY@it=\textit\def\PY@tc##1{\textcolor[rgb]{0.25,0.50,0.56}{##1}}}
\expandafter\def\csname PY@tok@vg\endcsname{\def\PY@tc##1{\textcolor[rgb]{0.73,0.38,0.84}{##1}}}
\expandafter\def\csname PY@tok@m\endcsname{\def\PY@tc##1{\textcolor[rgb]{0.13,0.50,0.31}{##1}}}
\expandafter\def\csname PY@tok@mh\endcsname{\def\PY@tc##1{\textcolor[rgb]{0.13,0.50,0.31}{##1}}}
\expandafter\def\csname PY@tok@cs\endcsname{\def\PY@tc##1{\textcolor[rgb]{0.25,0.50,0.56}{##1}}\def\PY@bc##1{\setlength{\fboxsep}{0pt}\colorbox[rgb]{1.00,0.94,0.94}{\strut ##1}}}
\expandafter\def\csname PY@tok@ge\endcsname{\let\PY@it=\textit}
\expandafter\def\csname PY@tok@vc\endcsname{\def\PY@tc##1{\textcolor[rgb]{0.73,0.38,0.84}{##1}}}
\expandafter\def\csname PY@tok@il\endcsname{\def\PY@tc##1{\textcolor[rgb]{0.13,0.50,0.31}{##1}}}
\expandafter\def\csname PY@tok@go\endcsname{\def\PY@tc##1{\textcolor[rgb]{0.20,0.20,0.20}{##1}}}
\expandafter\def\csname PY@tok@cp\endcsname{\def\PY@tc##1{\textcolor[rgb]{0.00,0.44,0.13}{##1}}}
\expandafter\def\csname PY@tok@gi\endcsname{\def\PY@tc##1{\textcolor[rgb]{0.00,0.63,0.00}{##1}}}
\expandafter\def\csname PY@tok@gh\endcsname{\let\PY@bf=\textbf\def\PY@tc##1{\textcolor[rgb]{0.00,0.00,0.50}{##1}}}
\expandafter\def\csname PY@tok@ni\endcsname{\let\PY@bf=\textbf\def\PY@tc##1{\textcolor[rgb]{0.84,0.33,0.22}{##1}}}
\expandafter\def\csname PY@tok@nl\endcsname{\let\PY@bf=\textbf\def\PY@tc##1{\textcolor[rgb]{0.00,0.13,0.44}{##1}}}
\expandafter\def\csname PY@tok@nn\endcsname{\let\PY@bf=\textbf\def\PY@tc##1{\textcolor[rgb]{0.05,0.52,0.71}{##1}}}
\expandafter\def\csname PY@tok@no\endcsname{\def\PY@tc##1{\textcolor[rgb]{0.38,0.68,0.84}{##1}}}
\expandafter\def\csname PY@tok@na\endcsname{\def\PY@tc##1{\textcolor[rgb]{0.25,0.44,0.63}{##1}}}
\expandafter\def\csname PY@tok@nb\endcsname{\def\PY@tc##1{\textcolor[rgb]{0.00,0.44,0.13}{##1}}}
\expandafter\def\csname PY@tok@nc\endcsname{\let\PY@bf=\textbf\def\PY@tc##1{\textcolor[rgb]{0.05,0.52,0.71}{##1}}}
\expandafter\def\csname PY@tok@nd\endcsname{\let\PY@bf=\textbf\def\PY@tc##1{\textcolor[rgb]{0.33,0.33,0.33}{##1}}}
\expandafter\def\csname PY@tok@ne\endcsname{\def\PY@tc##1{\textcolor[rgb]{0.00,0.44,0.13}{##1}}}
\expandafter\def\csname PY@tok@nf\endcsname{\def\PY@tc##1{\textcolor[rgb]{0.02,0.16,0.49}{##1}}}
\expandafter\def\csname PY@tok@si\endcsname{\let\PY@it=\textit\def\PY@tc##1{\textcolor[rgb]{0.44,0.63,0.82}{##1}}}
\expandafter\def\csname PY@tok@s2\endcsname{\def\PY@tc##1{\textcolor[rgb]{0.25,0.44,0.63}{##1}}}
\expandafter\def\csname PY@tok@vi\endcsname{\def\PY@tc##1{\textcolor[rgb]{0.73,0.38,0.84}{##1}}}
\expandafter\def\csname PY@tok@nt\endcsname{\let\PY@bf=\textbf\def\PY@tc##1{\textcolor[rgb]{0.02,0.16,0.45}{##1}}}
\expandafter\def\csname PY@tok@nv\endcsname{\def\PY@tc##1{\textcolor[rgb]{0.73,0.38,0.84}{##1}}}
\expandafter\def\csname PY@tok@s1\endcsname{\def\PY@tc##1{\textcolor[rgb]{0.25,0.44,0.63}{##1}}}
\expandafter\def\csname PY@tok@gp\endcsname{\let\PY@bf=\textbf\def\PY@tc##1{\textcolor[rgb]{0.78,0.36,0.04}{##1}}}
\expandafter\def\csname PY@tok@sh\endcsname{\def\PY@tc##1{\textcolor[rgb]{0.25,0.44,0.63}{##1}}}
\expandafter\def\csname PY@tok@ow\endcsname{\let\PY@bf=\textbf\def\PY@tc##1{\textcolor[rgb]{0.00,0.44,0.13}{##1}}}
\expandafter\def\csname PY@tok@sx\endcsname{\def\PY@tc##1{\textcolor[rgb]{0.78,0.36,0.04}{##1}}}
\expandafter\def\csname PY@tok@bp\endcsname{\def\PY@tc##1{\textcolor[rgb]{0.00,0.44,0.13}{##1}}}
\expandafter\def\csname PY@tok@c1\endcsname{\let\PY@it=\textit\def\PY@tc##1{\textcolor[rgb]{0.25,0.50,0.56}{##1}}}
\expandafter\def\csname PY@tok@kc\endcsname{\let\PY@bf=\textbf\def\PY@tc##1{\textcolor[rgb]{0.00,0.44,0.13}{##1}}}
\expandafter\def\csname PY@tok@c\endcsname{\let\PY@it=\textit\def\PY@tc##1{\textcolor[rgb]{0.25,0.50,0.56}{##1}}}
\expandafter\def\csname PY@tok@mf\endcsname{\def\PY@tc##1{\textcolor[rgb]{0.13,0.50,0.31}{##1}}}
\expandafter\def\csname PY@tok@err\endcsname{\def\PY@bc##1{\setlength{\fboxsep}{0pt}\fcolorbox[rgb]{1.00,0.00,0.00}{1,1,1}{\strut ##1}}}
\expandafter\def\csname PY@tok@kd\endcsname{\let\PY@bf=\textbf\def\PY@tc##1{\textcolor[rgb]{0.00,0.44,0.13}{##1}}}
\expandafter\def\csname PY@tok@ss\endcsname{\def\PY@tc##1{\textcolor[rgb]{0.32,0.47,0.09}{##1}}}
\expandafter\def\csname PY@tok@sr\endcsname{\def\PY@tc##1{\textcolor[rgb]{0.14,0.33,0.53}{##1}}}
\expandafter\def\csname PY@tok@mo\endcsname{\def\PY@tc##1{\textcolor[rgb]{0.13,0.50,0.31}{##1}}}
\expandafter\def\csname PY@tok@mi\endcsname{\def\PY@tc##1{\textcolor[rgb]{0.13,0.50,0.31}{##1}}}
\expandafter\def\csname PY@tok@kn\endcsname{\let\PY@bf=\textbf\def\PY@tc##1{\textcolor[rgb]{0.00,0.44,0.13}{##1}}}
\expandafter\def\csname PY@tok@o\endcsname{\def\PY@tc##1{\textcolor[rgb]{0.40,0.40,0.40}{##1}}}
\expandafter\def\csname PY@tok@kr\endcsname{\let\PY@bf=\textbf\def\PY@tc##1{\textcolor[rgb]{0.00,0.44,0.13}{##1}}}
\expandafter\def\csname PY@tok@s\endcsname{\def\PY@tc##1{\textcolor[rgb]{0.25,0.44,0.63}{##1}}}
\expandafter\def\csname PY@tok@kp\endcsname{\def\PY@tc##1{\textcolor[rgb]{0.00,0.44,0.13}{##1}}}
\expandafter\def\csname PY@tok@w\endcsname{\def\PY@tc##1{\textcolor[rgb]{0.73,0.73,0.73}{##1}}}
\expandafter\def\csname PY@tok@kt\endcsname{\def\PY@tc##1{\textcolor[rgb]{0.56,0.13,0.00}{##1}}}
\expandafter\def\csname PY@tok@sc\endcsname{\def\PY@tc##1{\textcolor[rgb]{0.25,0.44,0.63}{##1}}}
\expandafter\def\csname PY@tok@sb\endcsname{\def\PY@tc##1{\textcolor[rgb]{0.25,0.44,0.63}{##1}}}
\expandafter\def\csname PY@tok@k\endcsname{\let\PY@bf=\textbf\def\PY@tc##1{\textcolor[rgb]{0.00,0.44,0.13}{##1}}}
\expandafter\def\csname PY@tok@se\endcsname{\let\PY@bf=\textbf\def\PY@tc##1{\textcolor[rgb]{0.25,0.44,0.63}{##1}}}
\expandafter\def\csname PY@tok@sd\endcsname{\let\PY@it=\textit\def\PY@tc##1{\textcolor[rgb]{0.25,0.44,0.63}{##1}}}

\def\PYZus{\char`\_}
\def\PYZob{\char`\{}
\def\PYZcb{\char`\}}

\def\PYZsh{\char`\#}

\def\PYZhy{\char`\-}
\def\PYZsq{\char`\'}
\def\PYZdq{\char`\"}

\makeatother

\providecommand*{\DUfootnotemark}[3]{%
  \raisebox{1em}{\hypertarget{#1}{}}%
  \hyperlink{#2}{\textsuperscript{#3}}%
}
\providecommand{\DUfootnotetext}[4]{%
  \begingroup%
  \renewcommand{\thefootnote}{%
    \protect\raisebox{1em}{\protect\hypertarget{#1}{}}%
    \protect\hyperlink{#2}{#3}}%
  \footnotetext{#4}%
  \endgroup%
}

\providecommand*{\DUrole}[2]{%
  \ifcsname DUrole#1\endcsname%
    \csname DUrole#1\endcsname{#2}%
  \else
    \ifcsname docutilsrole#1\endcsname%
      \csname docutilsrole#1\endcsname{#2}%
    \else%
      #2%
    \fi%
  \fi%
}

\ifthenelse{\isundefined{\hypersetup}}{
  \usepackage[colorlinks=true,linkcolor=blue,urlcolor=blue]{hyperref}
  \urlstyle{same} 
}{}

\begin{document}
\newcounter{footnotecounter}\title{PySTEMM: Executable Concept Modeling for K-12 STEM Learning}\author{Kelsey D'Souza$^{\setcounter{footnotecounter}{1}\fnsymbol{footnotecounter}\setcounter{footnotecounter}{2}\fnsymbol{footnotecounter}}$%
          \setcounter{footnotecounter}{1}\thanks{\fnsymbol{footnotecounter} %
          Corresponding author: \protect\href{mailto:kelsey@dsouzaville.com}{kelsey@dsouzaville.com}}\setcounter{footnotecounter}{2}\thanks{\fnsymbol{footnotecounter} Senior at Westwood High School}\thanks{%

          \noindent%
          Copyright\,\copyright\,2014 Kelsey D'Souza. This is an open-access article distributed under the terms of the Creative Commons Attribution License, which permits unrestricted use, distribution, and reproduction in any medium, provided the original author and source are credited. http://creativecommons.org/licenses/by/3.0/%
        }}\maketitle
          \renewcommand{\leftmark}{PROC. OF THE 6th EUR. CONF. ON PYTHON IN SCIENCE (EUROSCIPY 2013)}
          \renewcommand{\rightmark}{PYSTEMM: EXECUTABLE CONCEPT MODELING FOR K-12 STEM LEARNING}

\setcounter{page}{37}
\newcommand*{\docutilsroleref}{\ref}
\newcommand*{\docutilsrolelabel}{\label}
\AtEndDocument{\cleardoublepage}




\begin{abstract}Modeling should play a central role in K-12 STEM education, where it could make classes much more engaging. A model underlies every scientific theory, and models are central to all the STEM disciplines (Science, Technology, Engineering, Math). This paper describes executable concept modeling of STEM concepts using immutable objects and pure functions in Python. I present examples in math, physics, chemistry, and engineering, built using a proof-of-concept tool called PySTEMM . The approach applies to all STEM areas and supports learning with pictures, narrative, animation, and graph plots. Models can extend each other, simplifying getting started. The functional-programming style reduces incidental complexity and code debugging.
\end{abstract}

\section{Introduction%
  \label{introduction}%
}

A \emph{model} is a simplified representation of part of some world, focused on selected aspects. A model underlies every scientific theory, and models are central to all STEM areas — science, technology, engineering, and mathematics — helping us conceptualize, understand, explain, and predict phenomena objectively. Children form mental models and physical models during play to understand their world. Scientists use bio-engineered tissue as a model of human organs. Computational modeling is revolutionizing science and engineering, as recognized by the 2013 Nobel Price in Chemistry going for computational modeling of biochemical systems.

Previous research \cite{Whi93}, \cite{Orn08} has shown significant learning benefits from model-building and exploring in STEM education. Students should create, validate, refute, and use models to better understand deep connections across subject areas, rather than mechanically drilling through problems. In this paper I demonstrate that executable concept modeling, based on using immutable objects and pure functions in Python:%
\begin{itemize}

\item 

applies across multiple STEM areas,
\item 

supports different representations and learning modes,
\item 

is feasible and approachable,
\item 

encourages bottom-up exploration and assembly, and
\item 

builds deep understanding of underlying concepts.
\end{itemize}


\subsection{Executable Concept Models%
  \label{executable-concept-models}%
}

A \emph{concept model} describes something by capturing relevant concepts, attributes, and rules. A \emph{concept instance} is a specific individual of a \emph{concept type} e.g. \texttt{NO2} is a concept instance of the general concept type \texttt{Molecule}. The concept type \texttt{Molecule} might define a \texttt{formula} attribute for any molecule instance to list how many atoms of each element it contains. The concept instance \texttt{NO2} has one \texttt{Nitrogen} and two \texttt{Oxygen} atoms. This is similar to the idea from object-oriented programming of an object that is an instance of a class.

Concepts and attributes are chosen to suit a purpose. A different model of \texttt{Molecule} might describe atoms, functional groups, bonds, sites at which other molecules can interact, site geometry, and forces that govern geometry and interactions.

An \emph{executable concept model} is represented on a computer, so concept instances and concept types can be manipulated and checked by the machine, increasing confidence in the model.

\subsection{PySTEMM Models%
  \label{pystemm-models}%
}


“Executable” typically entails programming language complexity, debugging headaches, and distractions from the actual concepts under study. Much of this complexity stems from \emph{imperative programming}, where variables and object attributes are modified as the program executes its procedures.

\emph{Functional programming} is a good alternative. It uses (a) \emph{immutable objects}, whose attribute values do not get modified by program code; and (b) \emph{pure functions}, producing a result that depends solely on inputs, without modifying any other attributes or variables.

PySTEMM, by using immutable objects and pure functions, and providing multiple model representations, reduces needless complexity and debugging. It uses the \emph{Python} programming language to define executable concept models that have three parts:\newcounter{listcnt0}
\begin{list}{\arabic{listcnt0}.}
{
\usecounter{listcnt0}
\setlength{\rightmargin}{\leftmargin}
}

\item 

Structure: A concept type is defined by a Python \emph{class} that describes attributes together with their types (which can reference other concept types). A concept instance is a Python \emph{object} instantiated from that class, with values for its attributes.
\item 

Functions: The pure functions that represent additional properties or rules on concept instances are defined as Python \emph{methods} on the class\DUfootnotemark{id3}{id4}{1}.
\item 

Visualization: The visualization of concept types and instances are defined with Python \emph{dictionaries} of visual properties, used as \emph{templates}.\end{list}


PySTEMM models focus on defining \emph{what terms and concepts mean}, rather than step-by-step instructions about \emph{how to compute}. PySTEMM functions manipulate not just numbers, but molecules, rigid bodies, planets, visualizations, and even concept types and functions.%
\DUfootnotetext{id4}{id3}{1}{
Since we use methods on a class for functions, in \textquotedbl{}\texttt{a.f(x)}\textquotedbl{} the inputs to \texttt{f} include argument \texttt{x}, and the object \texttt{a} on which the method is invoked.}

In the rest of this paper I present example models from math, chemistry, physics, and engineering, introduce key aspects of PySTEMM, and show  Python model source code as well as multiple model representations generated by PySTEMM. The last section describes the implementation of PySTEMM.

\section{Mathematics%
  \label{mathematics}%
}

We begin with models of math functions, because math forms the basis of all other models. Next we move on to \emph{high-order} functions i.e. functions that accept functions as inputs, or whose results are functions. Since our focus in this section is modeling math concepts, we will model math functions as objects. In subsequent sections on physics, chemistry, etc., we will directly use normal Python code for math computations.

\subsection{Basic Numeric Functions%
  \label{basic-numeric-functions}%
}
\begin{figure}[]\noindent\makebox[\columnwidth][c]{\includegraphics[width=\columnwidth]{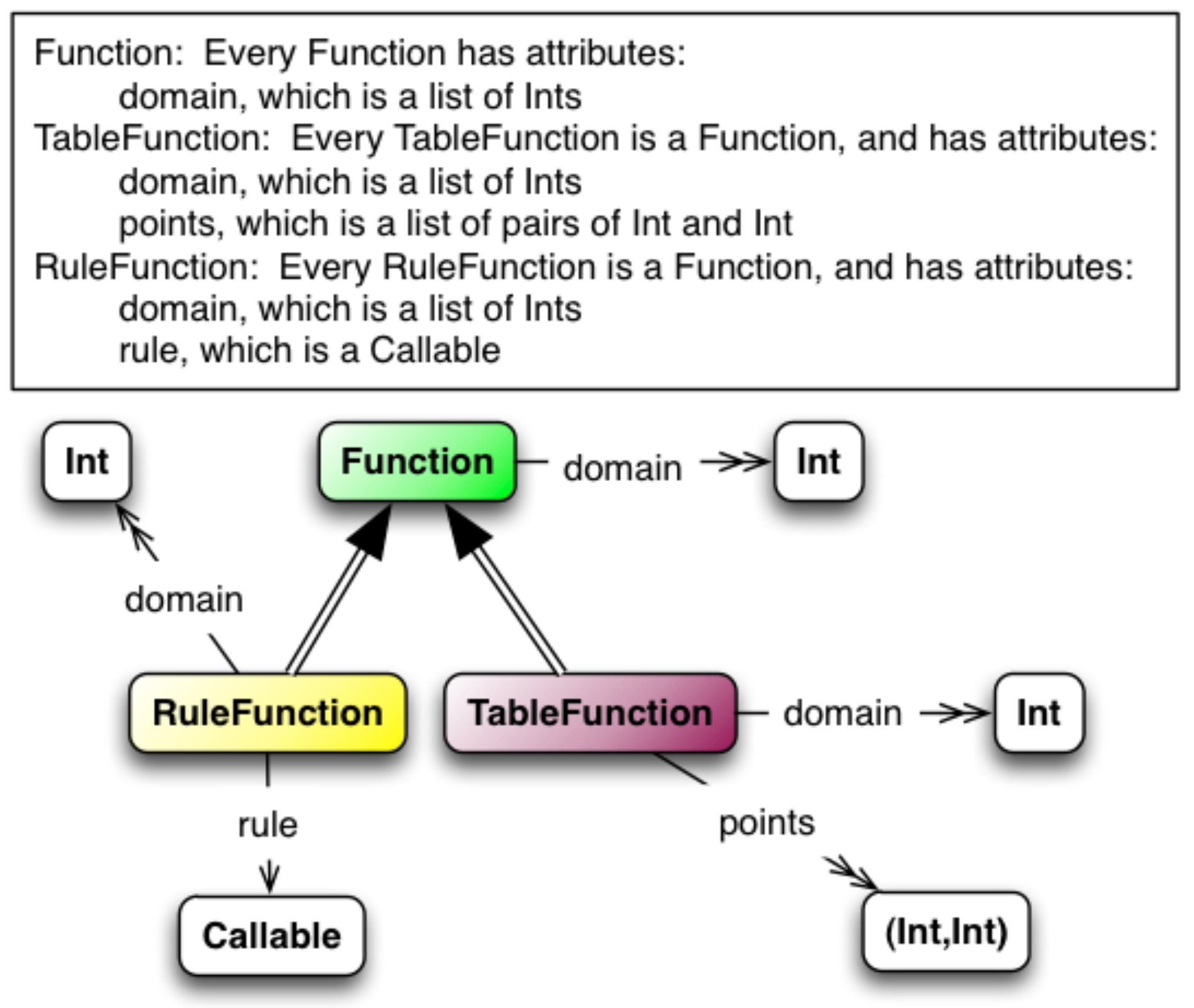}}
\caption{Three \texttt{Function} concept types. \DUrole{label}{functypes}}
\end{figure}

The Python model of \emph{concept types} for basic functions is:
\begin{Verbatim}[commandchars=\\\{\},numbers=left,firstnumber=1,stepnumber=1,fontsize=\footnotesize,xleftmargin=2.25mm,numbersep=3pt]
 \PY{c}{\PYZsh{} file: function\PYZus{}types.py}

 \PY{k}{class} \PY{n+nc}{Function}\PY{p}{(}\PY{n}{Concept}\PY{p}{)}\PY{p}{:}
   \PY{n}{domain} \PY{o}{=} \PY{n}{Property}\PY{p}{(}\PY{n}{List}\PY{p}{(}\PY{n}{Int}\PY{p}{)}\PY{p}{)}
   \PY{k}{def} \PY{n+nf}{eval}\PY{p}{(}\PY{n+nb+bp}{self}\PY{p}{,} \PY{n}{x}\PY{p}{)}\PY{p}{:} \PY{k}{pass}
   \PY{n}{class\PYZus{}template} \PY{o}{=} \PY{p}{\PYZob{}}\PY{n}{K}\PY{o}{.}\PY{n}{gradient\PYZus{}color}\PY{p}{:} \PY{l+s}{\PYZsq{}}\PY{l+s}{Green}\PY{l+s}{\PYZsq{}}\PY{p}{\PYZcb{}}

 \PY{k}{class} \PY{n+nc}{RuleFunction}\PY{p}{(}\PY{n}{Function}\PY{p}{)}\PY{p}{:}
   \PY{n}{rule} \PY{o}{=} \PY{n}{Callable}
   \PY{n}{domain} \PY{o}{=} \PY{n}{List}\PY{p}{(}\PY{n}{Int}\PY{p}{)}

   \PY{k}{def} \PY{n+nf}{eval}\PY{p}{(}\PY{n+nb+bp}{self}\PY{p}{,} \PY{n}{x}\PY{p}{)}\PY{p}{:}
     \PY{k}{return} \PY{n+nb+bp}{self}\PY{o}{.}\PY{n}{rule}\PY{p}{(}\PY{n}{x}\PY{p}{)}

   \PY{n}{class\PYZus{}template} \PY{o}{=} \PY{p}{\PYZob{}}\PY{n}{K}\PY{o}{.}\PY{n}{gradient\PYZus{}color}\PY{p}{:} \PY{l+s}{\PYZsq{}}\PY{l+s}{Yellow}\PY{l+s}{\PYZsq{}}\PY{p}{\PYZcb{}}

 \PY{k}{class} \PY{n+nc}{TableFunction}\PY{p}{(}\PY{n}{Function}\PY{p}{)}\PY{p}{:}
   \PY{n}{points} \PY{o}{=} \PY{n}{List}\PY{p}{(}\PY{n}{Tuple}\PY{p}{(}\PY{n}{Int}\PY{p}{,} \PY{n}{Int}\PY{p}{)}\PY{p}{)}
   \PY{n}{domain} \PY{o}{=} \PY{n}{Property}\PY{p}{(}\PY{n}{List}\PY{p}{(}\PY{n}{Int}\PY{p}{)}\PY{p}{)}

   \PY{k}{def} \PY{n+nf}{\PYZus{}get\PYZus{}domain}\PY{p}{(}\PY{n+nb+bp}{self}\PY{p}{)}\PY{p}{:}
     \PY{k}{return} \PY{p}{[}\PY{n}{x} \PY{k}{for} \PY{n}{x}\PY{p}{,} \PY{n}{y} \PY{o+ow}{in} \PY{n+nb+bp}{self}\PY{o}{.}\PY{n}{points}\PY{p}{]}

   \PY{k}{def} \PY{n+nf}{eval}\PY{p}{(}\PY{n+nb+bp}{self}\PY{p}{,} \PY{n}{x}\PY{p}{)}\PY{p}{:}
     \PY{k}{return} \PY{n}{find}\PY{p}{(}\PY{n}{y1} \PY{k}{for} \PY{n}{x1}\PY{p}{,}\PY{n}{y1} \PY{o+ow}{in} \PY{n+nb+bp}{self}\PY{o}{.}\PY{n}{points}
                   \PY{k}{if} \PY{n}{x1}\PY{o}{==}\PY{n}{x}\PY{p}{)}

   \PY{n}{class\PYZus{}template} \PY{o}{=} \PY{p}{\PYZob{}}\PY{n}{K}\PY{o}{.}\PY{n}{gradient\PYZus{}color}\PY{p}{:} \PY{l+s}{\PYZsq{}}\PY{l+s}{Maroon}\PY{l+s}{\PYZsq{}}\PY{p}{\PYZcb{}}
   \PY{n}{instance\PYZus{}template} \PY{o}{=} \PY{p}{\PYZob{}}\PY{n}{K}\PY{o}{.}\PY{n}{name}\PY{p}{:} \PY{l+s}{\PYZsq{}}\PY{l+s}{Circle}\PY{l+s}{\PYZsq{}}\PY{p}{\PYZcb{}}
\end{Verbatim}
The concept type \texttt{Function} is defined as a class (line 3), with an attribute \texttt{domain} which is a list of integers (line 4). \textquotedbl{}\texttt{Property}\textquotedbl{} allows \texttt{domain} to be represented differently for different subclasses of \texttt{Function}. Function evaluation is modeled by method \texttt{eval} (line 5) whose specifics are deferred to subclasses. The visualization of functions is defined by \texttt{class\_template} (line 6).

We define two subclasses of \texttt{Function}, each with different representations. \texttt{RuleFunctions} (line 8-15) are defined by an attribute \texttt{rule} that is a Python \emph{callable} expression, an explicit \texttt{domain}, and  \texttt{eval} that simply invokes \texttt{rule}. \texttt{TableFunctions} (line 17-29) are defined by a list of \texttt{(x,y)} pairs in an attribute \texttt{points}, a \texttt{domain}  computed from \texttt{points} by \texttt{\_get\_domain}, and \texttt{eval} that finds the matching pair in \texttt{points}. The \texttt{class\_template} (lines 15, 28) is a dictionary of visualization properties for the concept type, and \texttt{instance\_template} (line 29) is for visualizing instances. PySTEMM generates the visual and English narrative in Figure \DUrole{ref}{functypes} for  these concept types.\begin{figure}[]\noindent\makebox[\columnwidth][c]{\includegraphics[width=\columnwidth]{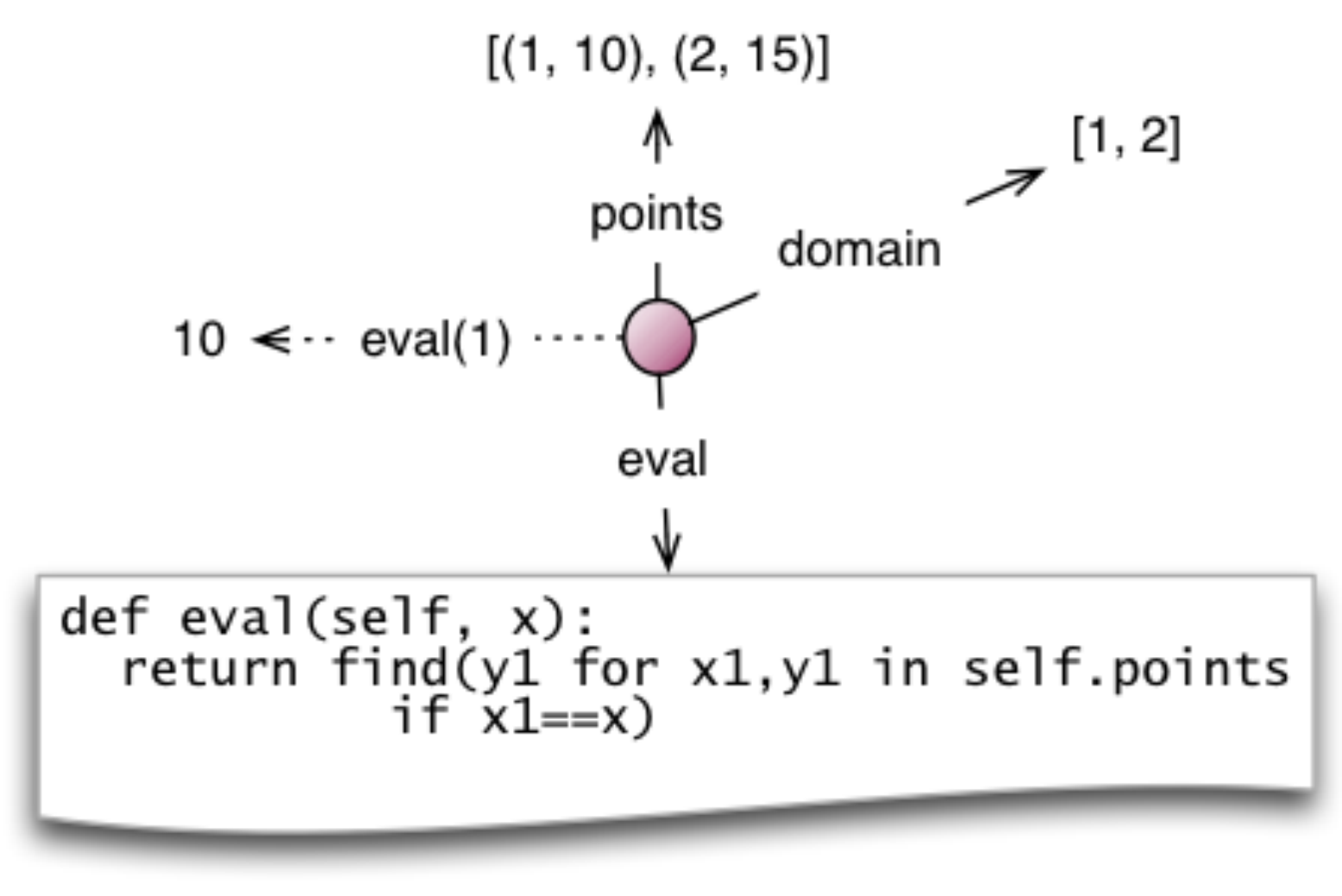}}
\caption{\texttt{TableFunction} concept instance. \DUrole{label}{funcinstances}}
\end{figure}

Below, we \emph{extend} this model with a \texttt{TableFunction} instance \texttt{tf} with its list of \texttt{points} (line 4), and customize what the model should visualize:\begin{Verbatim}[commandchars=\\\{\},numbers=left,firstnumber=1,stepnumber=1,fontsize=\footnotesize,xleftmargin=2.25mm,numbersep=3pt]
 \PY{c}{\PYZsh{} file function\PYZus{}instances.py}
 \PY{k+kn}{from} \PY{n+nn}{function\PYZus{}types.py} \PY{k+kn}{import} \PY{o}{*}

 \PY{n}{tf} \PY{o}{=} \PY{n}{TableFunction}\PY{p}{(}\PY{n}{points}\PY{o}{=}\PY{p}{[}\PY{p}{(}\PY{l+m+mi}{1}\PY{p}{,} \PY{l+m+mi}{10}\PY{p}{)}\PY{p}{,} \PY{p}{(}\PY{l+m+mi}{2}\PY{p}{,} \PY{l+m+mi}{15}\PY{p}{)}\PY{p}{]}\PY{p}{)}

 \PY{n}{M} \PY{o}{=} \PY{n}{Model}\PY{p}{(}\PY{p}{)}
 \PY{n}{M}\PY{o}{.}\PY{n}{addInstances}\PY{p}{(}\PY{n}{tf}\PY{p}{)}
 \PY{n}{M}\PY{o}{.}\PY{n}{showMethod}\PY{p}{(}\PY{n}{tf}\PY{p}{,} \PY{l+s}{\PYZsq{}}\PY{l+s}{eval}\PY{l+s}{\PYZsq{}}\PY{p}{)}
 \PY{n}{M}\PY{o}{.}\PY{n}{showEval}\PY{p}{(}\PY{n}{tf}\PY{p}{,}\PY{l+s}{\PYZsq{}}\PY{l+s}{eval}\PY{l+s}{\PYZsq{}}\PY{p}{,}\PY{p}{[}\PY{l+m+mi}{1}\PY{p}{]}\PY{p}{)}
\end{Verbatim}

PySTEMM generates  the visualization in Figure \DUrole{ref}{funcinstances}. The \texttt{domain} of \texttt{tf} was calculated from its \texttt{points}, its value at \texttt{x=1} is \texttt{10}, and the code for \texttt{eval()} is shown in the context of the instance. Since \texttt{eval} is a \emph{pure function}, \texttt{tf.eval(1)} depends solely on the input \texttt{1} and the definition of \texttt{tf} itself, so it is easy to understand the source code: it returns the \texttt{y1} from the \texttt{x1,y1} pair that matches the input \texttt{x}.

Note that \texttt{tf} is drawn as a circle of the same color as the \texttt{TableFunction} class: the \texttt{instance\_template} for \texttt{TableFunction} is merged with the \texttt{class\_template} before being applied to \texttt{tf}.

\subsection{Inverse Functions%
  \label{inverse-functions}%
}
\begin{figure}[]\noindent\makebox[\columnwidth][c]{\includegraphics[width=\columnwidth]{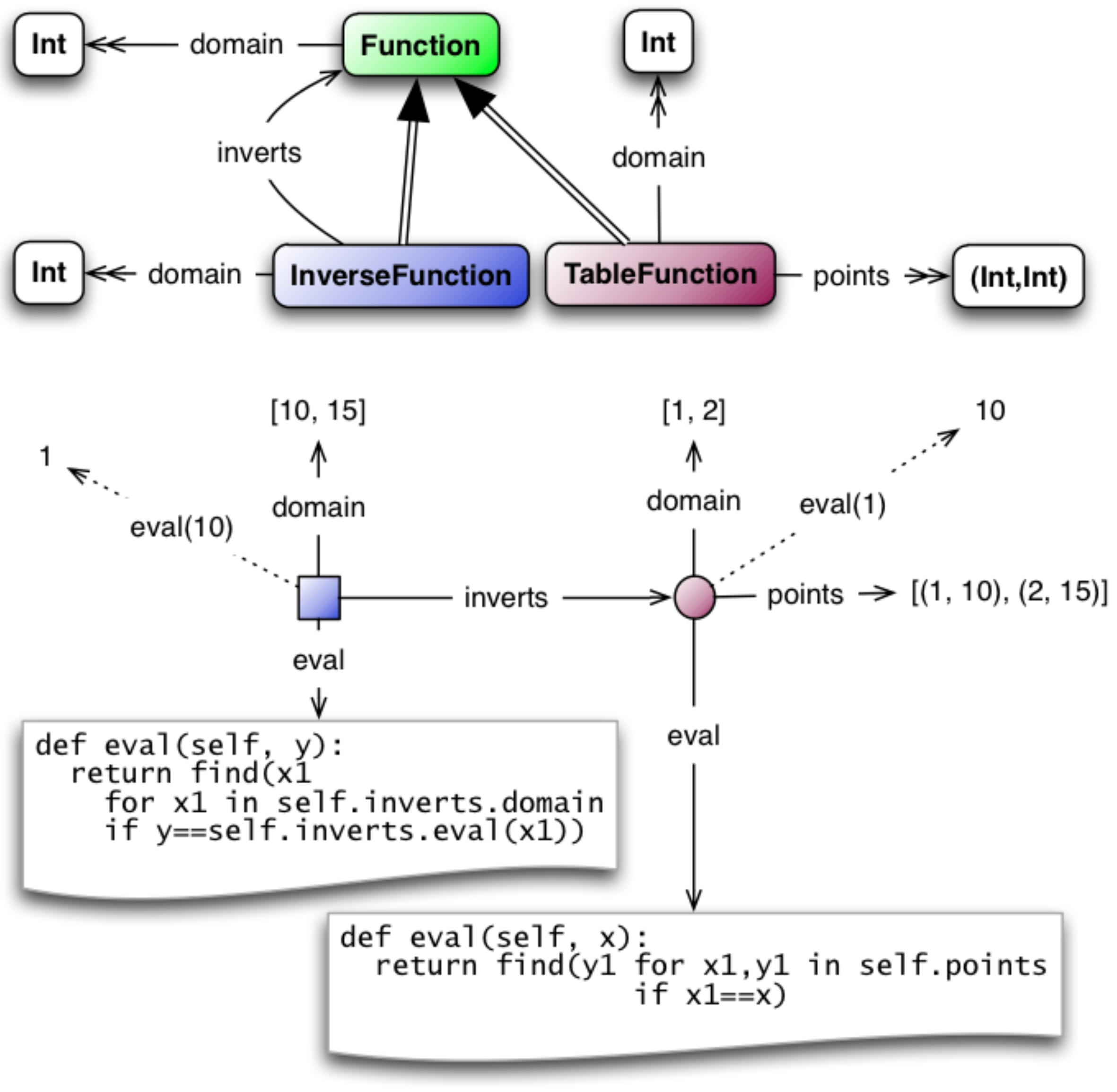}}
\caption{\texttt{InverseFunction} type and instance. \DUrole{label}{funcinverse}}
\end{figure}

An \texttt{InverseFunction} inverts another: $g = f^{-1}(x)$. The model below extends the \texttt{function\_instances} model with a class and an instance. On line 5, the \texttt{InverseFunction(...)} constructor is a \emph{high-order function} corresponding to the $f^{-1}$ operator, since it receives a function \texttt{tf} to invert, and produces the new inverted function \texttt{inv}.\begin{Verbatim}[commandchars=\\\{\},numbers=left,firstnumber=1,stepnumber=1,fontsize=\footnotesize,xleftmargin=2.25mm,numbersep=3pt]
\PY{k+kn}{from} \PY{n+nn}{function\PYZus{}instances} \PY{k+kn}{import} \PY{o}{*}

\PY{k}{class} \PY{n+nc}{InverseFunction}\PY{p}{(}\PY{n}{Concept}\PY{p}{)}\PY{p}{:} \PY{o}{.}\PY{o}{.}\PY{o}{.}

\PY{n}{inv} \PY{o}{=} \PY{n}{InverseFunction}\PY{p}{(}\PY{n}{inverts}\PY{o}{=}\PY{n}{tf}\PY{p}{)}

\PY{n}{M}\PY{o}{.}\PY{n}{addClasses}\PY{p}{(}\PY{n}{InverseFunction}\PY{p}{)}
\PY{n}{M}\PY{o}{.}\PY{n}{addInstances}\PY{p}{(}\PY{n}{inv}\PY{p}{)}
\PY{n}{M}\PY{o}{.}\PY{n}{showEval}\PY{p}{(}\PY{n}{inv}\PY{p}{,} \PY{l+s}{\PYZsq{}}\PY{l+s}{eval}\PY{l+s}{\PYZsq{}}\PY{p}{,}\PY{p}{[}\PY{l+m+mi}{15}\PY{p}{]}\PY{p}{)}
\end{Verbatim}
The instance visualization generated by PySTEMM in Figure \DUrole{ref}{funcinverse} shows the inverse function as a blue square, its \texttt{eval()} effectively flips the \texttt{(x,y)} pairs of the function it inverts, and its \texttt{domain} is computed as the set of \texttt{y} values of the function it inverts.

\subsection{Graph Transforms and High-Order Functions%
  \label{graph-transforms-and-high-order-functions}%
}
\begin{figure*}[]\noindent\makebox[\textwidth][c]{\includegraphics[scale=0.40]{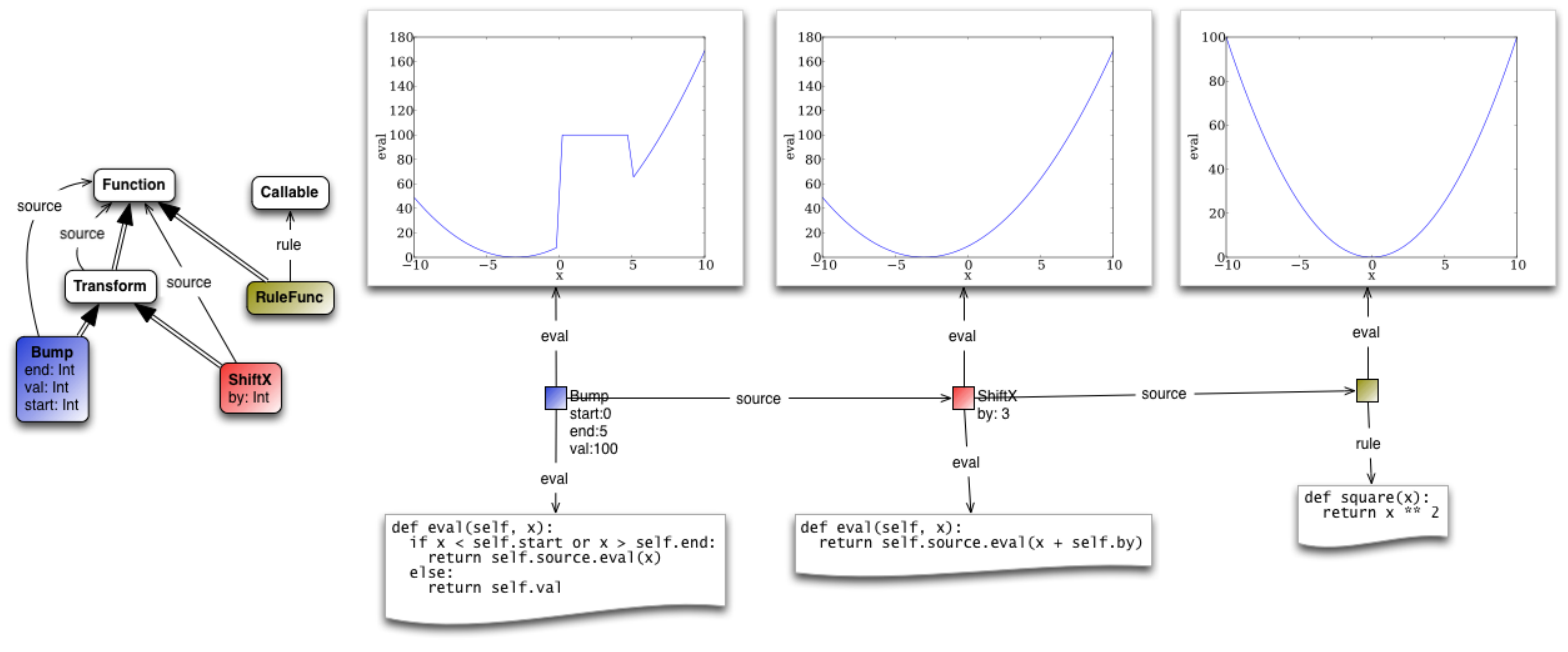}}
\caption{Function Transforms: A \texttt{Bump} of a \texttt{Shift} of $x^{2}$. \DUrole{label}{funcbump}}
\end{figure*}

A graph transformation as taught in middle school — translation, scaling,  rotation — is modeled as a function that operates on a \texttt{source} function, producing the transformed function. In Figure \DUrole{ref}{funcbump}, PySTEMM generates a graph plot of the original function, a shifted version, and a “bumped” version of the shifted function. The instances are defined as:
\begin{Verbatim}[commandchars=\\\{\},fontsize=\footnotesize]
\PY{n}{Bump}\PY{p}{(}\PY{n}{source} \PY{o}{=}
        \PY{n}{ShiftX}\PY{p}{(}\PY{n}{source} \PY{o}{=} \PY{n}{RuleFunc}\PY{p}{(}\PY{n}{rule}\PY{o}{=}\PY{n}{square}\PY{p}{)}\PY{p}{,}
               \PY{n}{by}\PY{o}{=}\PY{l+m+mi}{3}\PY{p}{)}\PY{p}{,}
     \PY{n}{start}\PY{o}{=}\PY{l+m+mi}{0}\PY{p}{,} \PY{n}{end}\PY{o}{=}\PY{l+m+mi}{5}\PY{p}{,} \PY{n}{val}\PY{o}{=}\PY{l+m+mi}{100}\PY{p}{)}
\end{Verbatim}
Similarly, the \emph{limit} of a function is a high-order function: it operates on another function and a target point, and evaluates to a single numeric value. Calculus operators, such as \emph{differentiation} and \emph{integration}, can be modeled as high-order functions as well: they operate on a function and produce a new function.


\section{Chemistry: Reaction%
  \label{chemistry-reaction}%
}
\begin{figure}[]\noindent\makebox[\columnwidth][c]{\includegraphics[width=\columnwidth]{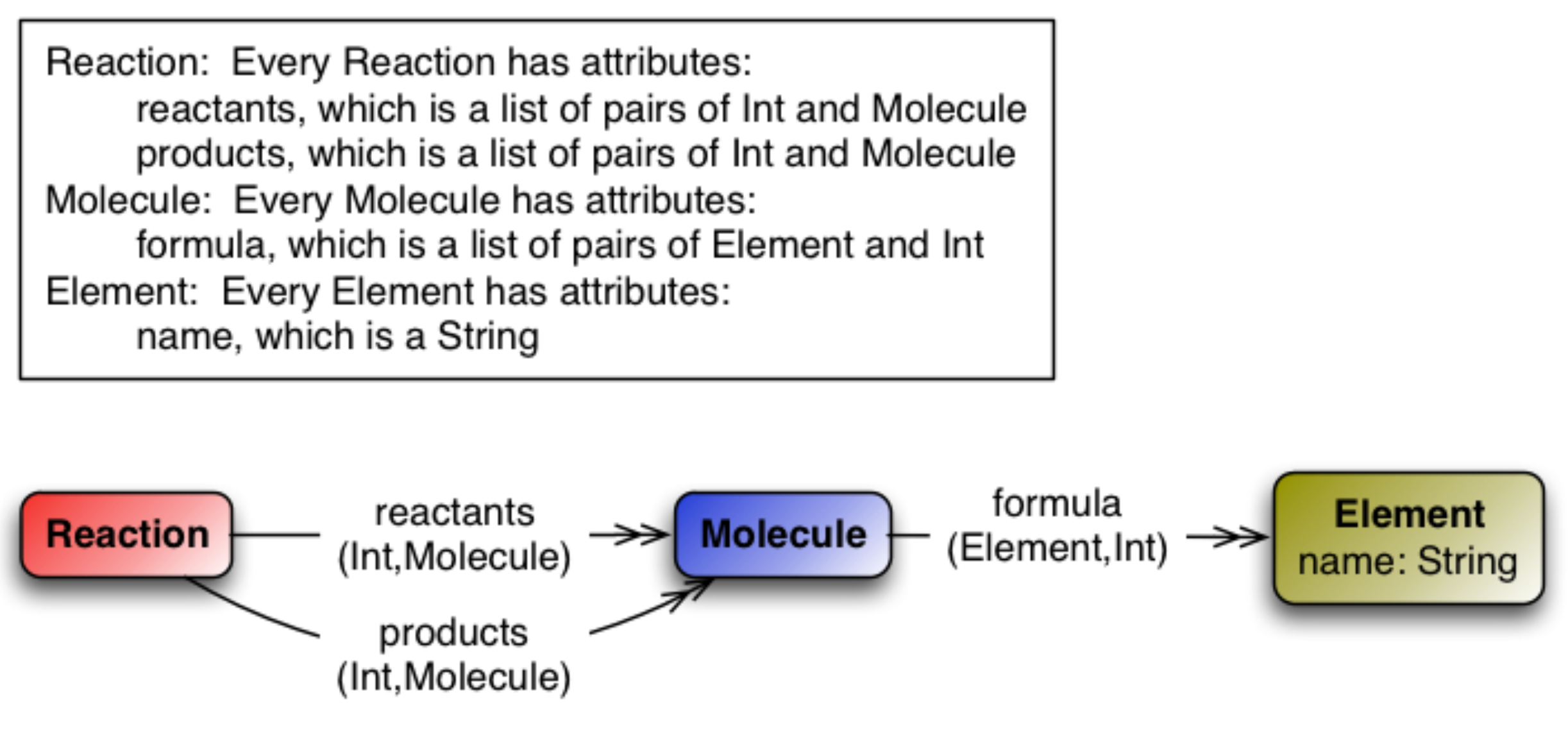}}
\caption{\texttt{Reaction} concept type. \DUrole{label}{reactiontypes}}
\end{figure}\begin{figure}[]\noindent\makebox[\columnwidth][c]{\includegraphics[width=\columnwidth]{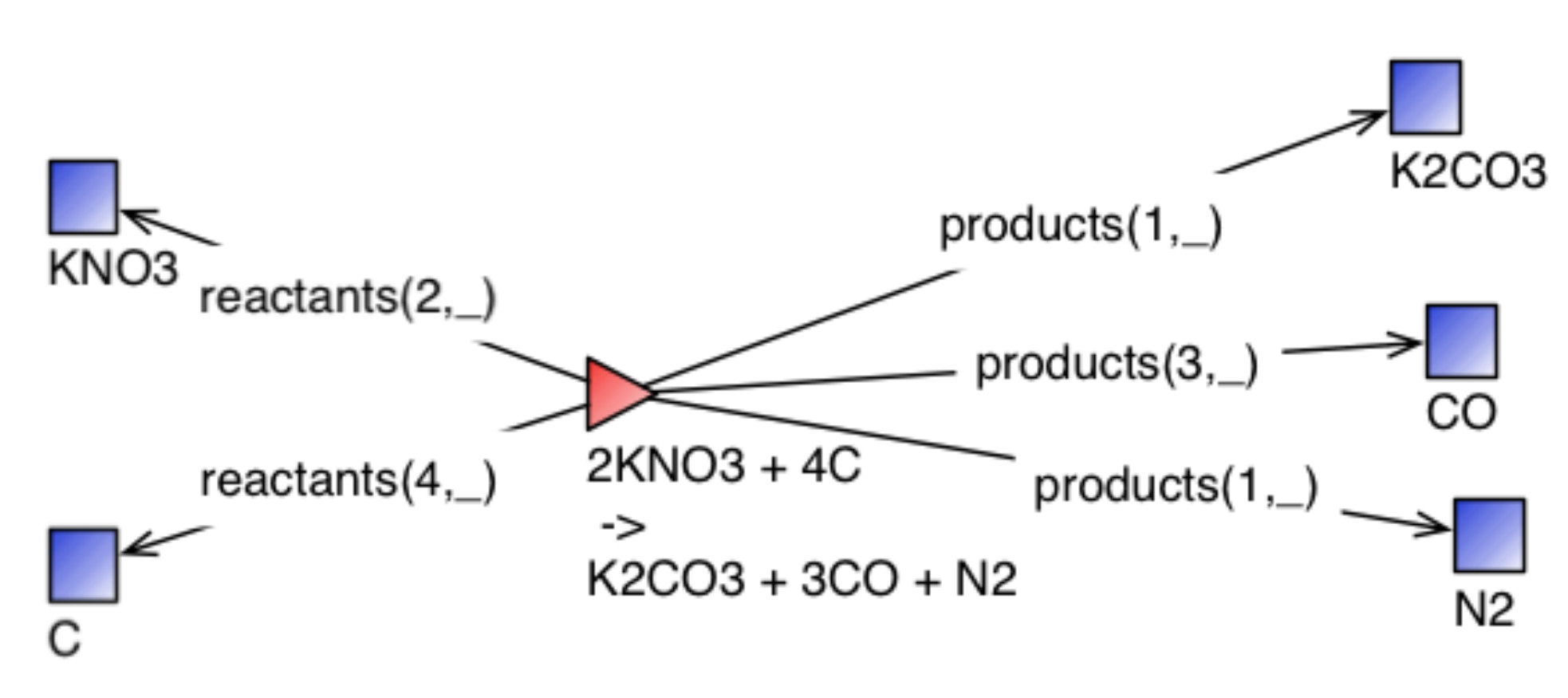}}
\caption{An instance of \texttt{Reaction}. \DUrole{label}{reactioninstance}}
\end{figure}\begin{Verbatim}[commandchars=\\\{\},numbers=left,firstnumber=1,stepnumber=1,fontsize=\footnotesize,xleftmargin=2.25mm,numbersep=3pt]
\PY{k}{class} \PY{n+nc}{Element}\PY{p}{(}\PY{n}{Concept}\PY{p}{)}\PY{p}{:}
  \PY{n}{name} \PY{o}{=} \PY{n}{String}

\PY{k}{class} \PY{n+nc}{Molecule}\PY{p}{(}\PY{n}{Concept}\PY{p}{)}\PY{p}{:}
  \PY{n}{formula} \PY{o}{=} \PY{n}{List}\PY{p}{(}\PY{n}{Tuple}\PY{p}{(}\PY{n}{Element}\PY{p}{,} \PY{n}{Int}\PY{p}{)}\PY{p}{)}
  \PY{n}{instance\PYZus{}template} \PY{o}{=} \PY{p}{\PYZob{}}
    \PY{n}{K}\PY{o}{.}\PY{n}{text}\PY{p}{:} \PY{k}{lambda} \PY{n}{m}\PY{p}{:} \PY{n}{computed\PYZus{}label}\PY{p}{(}\PY{n}{m}\PY{p}{)}\PY{p}{\PYZcb{}}

\PY{k}{class} \PY{n+nc}{Reaction}\PY{p}{(}\PY{n}{Concept}\PY{p}{)}\PY{p}{:}
  \PY{n}{products} \PY{o}{=} \PY{n}{List}\PY{p}{(}\PY{n}{Tuple}\PY{p}{(}\PY{n}{Int}\PY{p}{,} \PY{n}{Molecule}\PY{p}{)}\PY{p}{)}
  \PY{n}{reactants} \PY{o}{=} \PY{n}{List}\PY{p}{(}\PY{n}{Tuple}\PY{p}{(}\PY{n}{Int}\PY{p}{,} \PY{n}{Molecule}\PY{p}{)}\PY{p}{)}
\end{Verbatim}
An \texttt{Element} is modeled as just a name, since we ignore electron and nuclear structure. A \texttt{Molecule} has an attribute \texttt{formula} with a list of pairs of element with a number indicating the number of atoms of that element. A \texttt{Reaction} has \texttt{reactants} and \texttt{products}, each some quantity of a certain molecule. This Python model is visualized by PySTEMM in Figure \DUrole{ref}{reactiontypes}.

Note that convenient Python constructs, like \emph{lists} of \emph{tuples}, are visualized in a similarly convenient manner. Also, the \texttt{instance\_template} for molecule (lines 6-7), specifying the visualization properties for a molecule instance, contains a \emph{function} which takes a molecule instance and computes its label. Visualization templates are parameterized by the objects they will be applied to.

Figure \DUrole{ref}{reactioninstance} shows an instance of a reaction, showing reaction structure and computed labels for reactions and molecules, while hiding the \texttt{formula} structure within molecules.

\subsection{Reaction Balancing%
  \label{reaction-balancing}%
}
\begin{figure}[]\noindent\makebox[\columnwidth][c]{\includegraphics[width=\columnwidth]{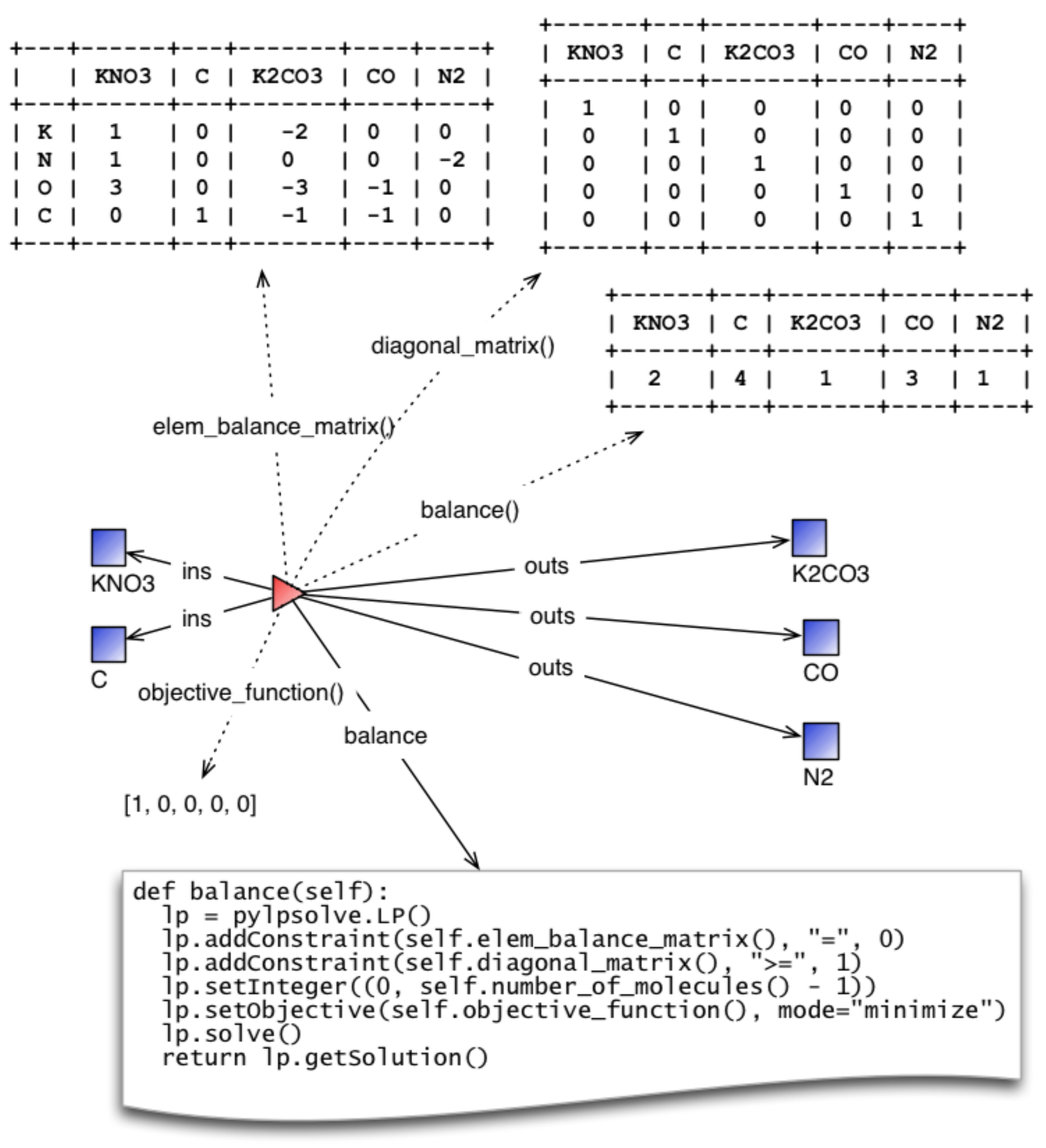}}
\caption{\texttt{Reaction} balance matrix and solved coefficients. \DUrole{label}{balancing}}
\end{figure}

Our next model computes reaction balancing for reactions. An unbalanced reaction has lists \texttt{ins} and \texttt{outs} of  molecules without coefficients. Figure \DUrole{ref}{balancing} shows how PySTEMM visualizes a reaction with the \texttt{balance} computation, coefficients, and intermediate values, as explained below.


We formulate reaction-balancing as an \emph{integer-linear programming} problem \cite{Sen06}, which we solve for molecule coefficients. The \texttt{formula} of the  molecules constrain the coefficients, since atoms of every element must balance. The function \texttt{elem\_balance\_matrix} computes a matrix of \emph{molecule} vs. \emph{element}, with the number of atoms of each element in each molecule, with \texttt{+} for \texttt{ins} and \texttt{-} for \texttt{outs}. This matrix multiplied by the vector of coefficients must result in all \texttt{0}. All coefficients have to be positive integers (\texttt{diagonal\_matrix}), and the \texttt{objective\_function} seeks the smallest coefficients  satisfying these constraints.

Once we have balanced reactions, we can add attributes and functions to model reaction stoichiometry and thermodynamics. For example:\begin{Verbatim}[commandchars=\\\{\},fontsize=\footnotesize]
\PY{k}{class} \PY{n+nc}{Element}\PY{p}{(}\PY{n}{Concept}\PY{p}{)}\PY{p}{:}
  \PY{n}{name} \PY{o}{=} \PY{n}{String}
  \PY{n}{atomic\PYZus{}mass} \PY{o}{=} \PY{n}{Float}

\PY{k}{class} \PY{n+nc}{Molecule}\PY{p}{(}\PY{n}{Concept}\PY{p}{)}\PY{p}{:}
  \PY{n}{formula} \PY{o}{=} \PY{n}{List}\PY{p}{(}\PY{n}{Tuple}\PY{p}{(}\PY{n}{Element}\PY{p}{,} \PY{n}{Int}\PY{p}{)}\PY{p}{)}
  \PY{n}{molar\PYZus{}mass} \PY{o}{=} \PY{n}{Property}\PY{p}{(}\PY{n}{Float}\PY{p}{)}
  \PY{k}{def} \PY{n+nf}{\PYZus{}get\PYZus{}molar\PYZus{}mass}\PY{p}{(}\PY{n+nb+bp}{self}\PY{p}{)}\PY{p}{:}
    \PY{k}{return} \PY{n+nb}{sum}\PY{p}{(}\PY{p}{[}\PY{n}{n} \PY{o}{*} \PY{n}{el}\PY{o}{.}\PY{n}{atomic\PYZus{}mass}
                  \PY{k}{for} \PY{n}{el}\PY{p}{,} \PY{n}{n} \PY{o+ow}{in} \PY{n+nb+bp}{self}\PY{o}{.}\PY{n}{formula}\PY{p}{]}\PY{p}{)}

\PY{n}{Fe} \PY{o}{=} \PY{n}{Element}\PY{p}{(}\PY{n}{name}\PY{o}{=}\PY{l+s}{\PYZsq{}}\PY{l+s}{Fe}\PY{l+s}{\PYZsq{}}\PY{p}{,} \PY{n}{atomic\PYZus{}mass}\PY{o}{=}\PY{l+m+mi}{56}\PY{p}{)}
\PY{n}{Cl} \PY{o}{=} \PY{n}{Element}\PY{p}{(}\PY{n}{name}\PY{o}{=}\PY{l+s}{\PYZsq{}}\PY{l+s}{Cl}\PY{l+s}{\PYZsq{}}\PY{p}{,} \PY{n}{atomic\PYZus{}mass}\PY{o}{=}\PY{l+m+mf}{35.5}\PY{p}{)}
\PY{n}{FeCl2} \PY{o}{=} \PY{n}{Molecule}\PY{p}{(}\PY{n}{formula}\PY{o}{=}\PY{p}{[}\PY{p}{(}\PY{n}{Fe}\PY{p}{,}\PY{l+m+mi}{1}\PY{p}{)}\PY{p}{,} \PY{p}{(}\PY{n}{Cl}\PY{p}{,}\PY{l+m+mi}{2}\PY{p}{)}\PY{p}{]}\PY{p}{)}

\PY{n}{FeCl2}\PY{o}{.}\PY{n}{molar\PYZus{}mass} \PY{c}{\PYZsh{} = 127}
\end{Verbatim}


\subsection{Reaction Network%
  \label{reaction-network}%
}
\begin{Verbatim}[commandchars=\\\{\},fontsize=\footnotesize]
\PY{k}{class} \PY{n+nc}{Network}\PY{p}{(}\PY{n}{Concept}\PY{p}{)}\PY{p}{:}
  \PY{n}{reactions} \PY{o}{=} \PY{n}{List}\PY{p}{(}\PY{n}{Reaction}\PY{p}{)}

\PY{n}{R1} \PY{o}{=} \PY{n}{Reaction}\PY{p}{(}\PY{n}{reactants}\PY{o}{=}\PY{p}{[}\PY{p}{(}\PY{l+m+mi}{2}\PY{p}{,} \PY{n}{NO2}\PY{p}{)}\PY{p}{]}\PY{p}{,}
              \PY{n}{products}\PY{o}{=}\PY{p}{[}\PY{p}{(}\PY{l+m+mi}{1}\PY{p}{,} \PY{n}{NO3}\PY{p}{)}\PY{p}{,} \PY{p}{(}\PY{l+m+mi}{1}\PY{p}{,} \PY{n}{NO}\PY{p}{)}\PY{p}{]}\PY{p}{)}

\PY{n}{R2} \PY{o}{=} \PY{n}{Reaction}\PY{p}{(}\PY{n}{reactants}\PY{o}{=}\PY{p}{[}\PY{p}{(}\PY{l+m+mi}{1}\PY{p}{,} \PY{n}{NO3}\PY{p}{)}\PY{p}{,} \PY{p}{(}\PY{l+m+mi}{1}\PY{p}{,} \PY{n}{CO}\PY{p}{)}\PY{p}{]}\PY{p}{,}
              \PY{n}{products}\PY{o}{=}\PY{p}{[}\PY{p}{(}\PY{l+m+mi}{1}\PY{p}{,} \PY{n}{NO2}\PY{p}{)}\PY{p}{,} \PY{p}{(}\PY{l+m+mi}{1}\PY{p}{,} \PY{n}{CO2}\PY{p}{)}\PY{p}{]}\PY{p}{)}

\PY{n}{Net} \PY{o}{=} \PY{n}{Network}\PY{p}{(}\PY{n}{reactions}\PY{o}{=}\PY{p}{[}\PY{n}{R1}\PY{p}{,} \PY{n}{R2}\PY{p}{]}\PY{p}{)}
\end{Verbatim}
\begin{figure}[]\noindent\makebox[\columnwidth][c]{\includegraphics[width=\columnwidth]{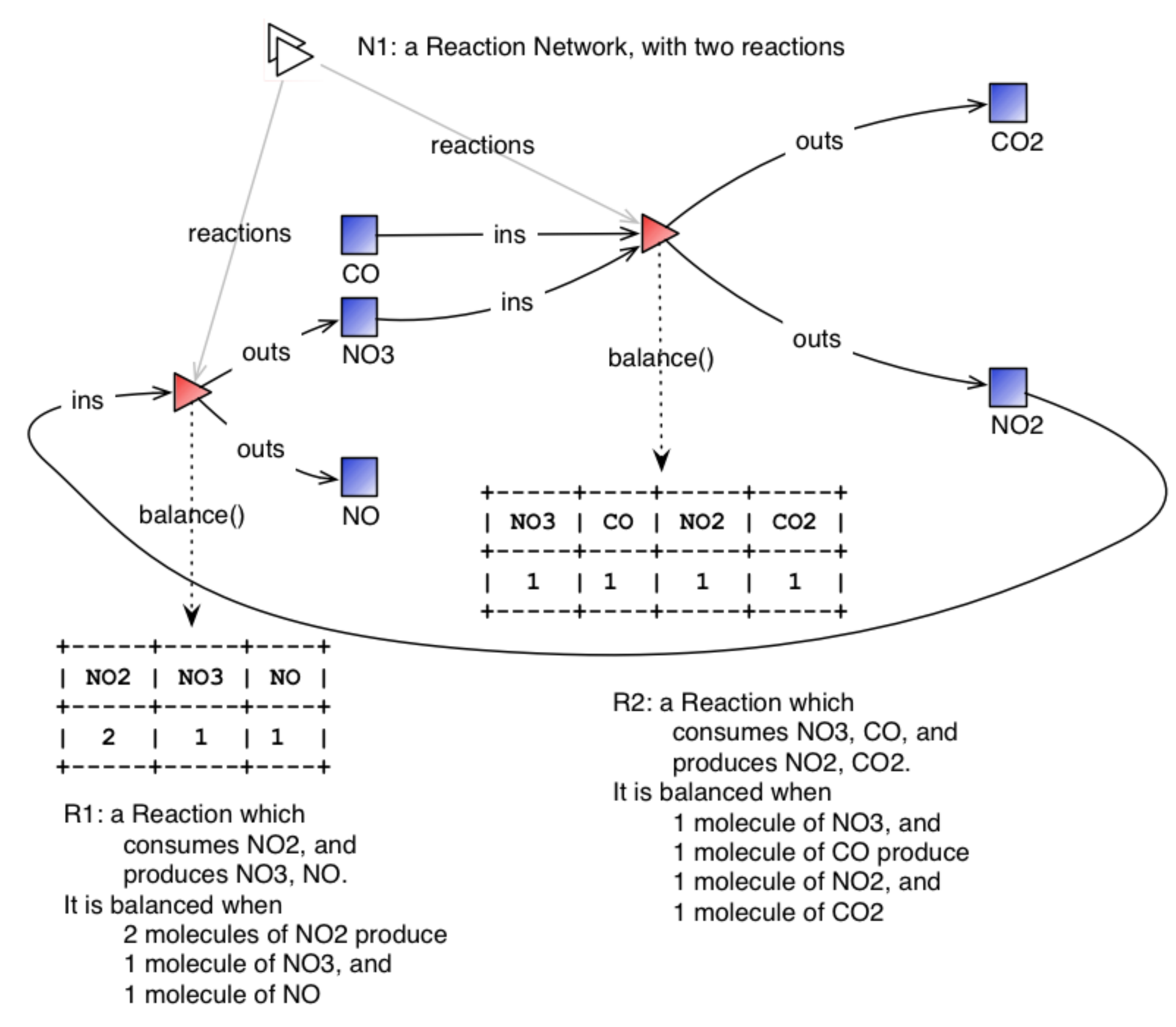}}
\caption{A reaction \texttt{Network} with two reactions. \DUrole{label}{network}}
\end{figure}A \texttt{Network} of coupled chemical reactions has a list of \texttt{reactions}. Given this Python model, and a narrative template for \texttt{Reaction}, PySTEMM generates Figure \DUrole{ref}{network}, including the \emph{instance-level} English narrative. Just as there are element balance constraints on an individual reaction, we could model network-level constraints on the reaction rates and concentrations of chemical species, but have not shown this here.

\subsection{Layered Models%
  \label{layered-models}%
}
\begin{figure}[]\noindent\makebox[\columnwidth][c]{\includegraphics[scale=0.65]{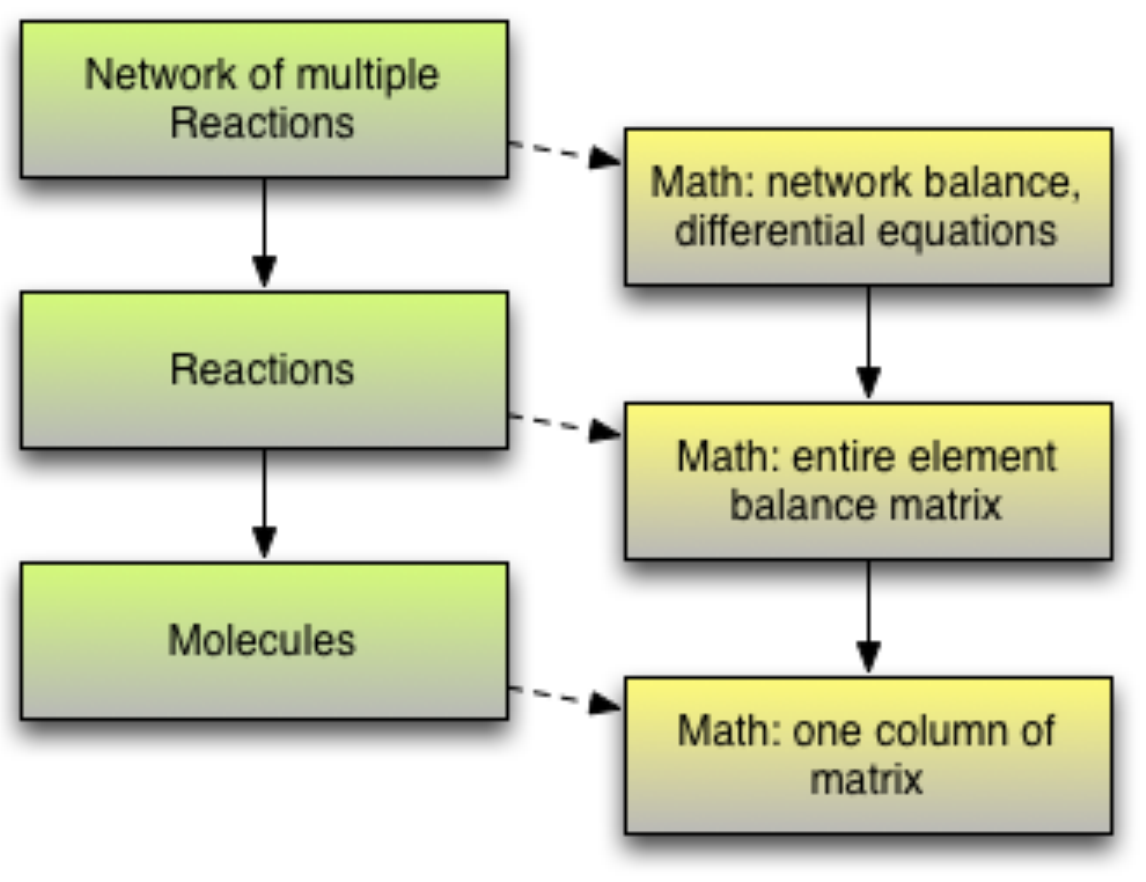}}
\caption{Layered concept models and generated math.}
\end{figure}

The reaction examples illustrate an important advantage of PySTEMM  modeling; instead of directly modeling the mathematics of reaction, we focus on the structure of the concept instances; in this case, what constitutes a molecule, or a reaction?

From this model, we compute the math model. The math version of a molecule is a single column with the number of atoms of each element type in that molecule. The math for a reaction collects this column from each molecule and combines them into an \texttt{element\_balance\_matrix}. Pure functions thus  easily traverse the concept instances to build corresponding math models such as matrices of numbers.

\section{Physics%
  \label{physics}%
}
\begin{figure*}[]\noindent\makebox[\textwidth][c]{\includegraphics[scale=0.40]{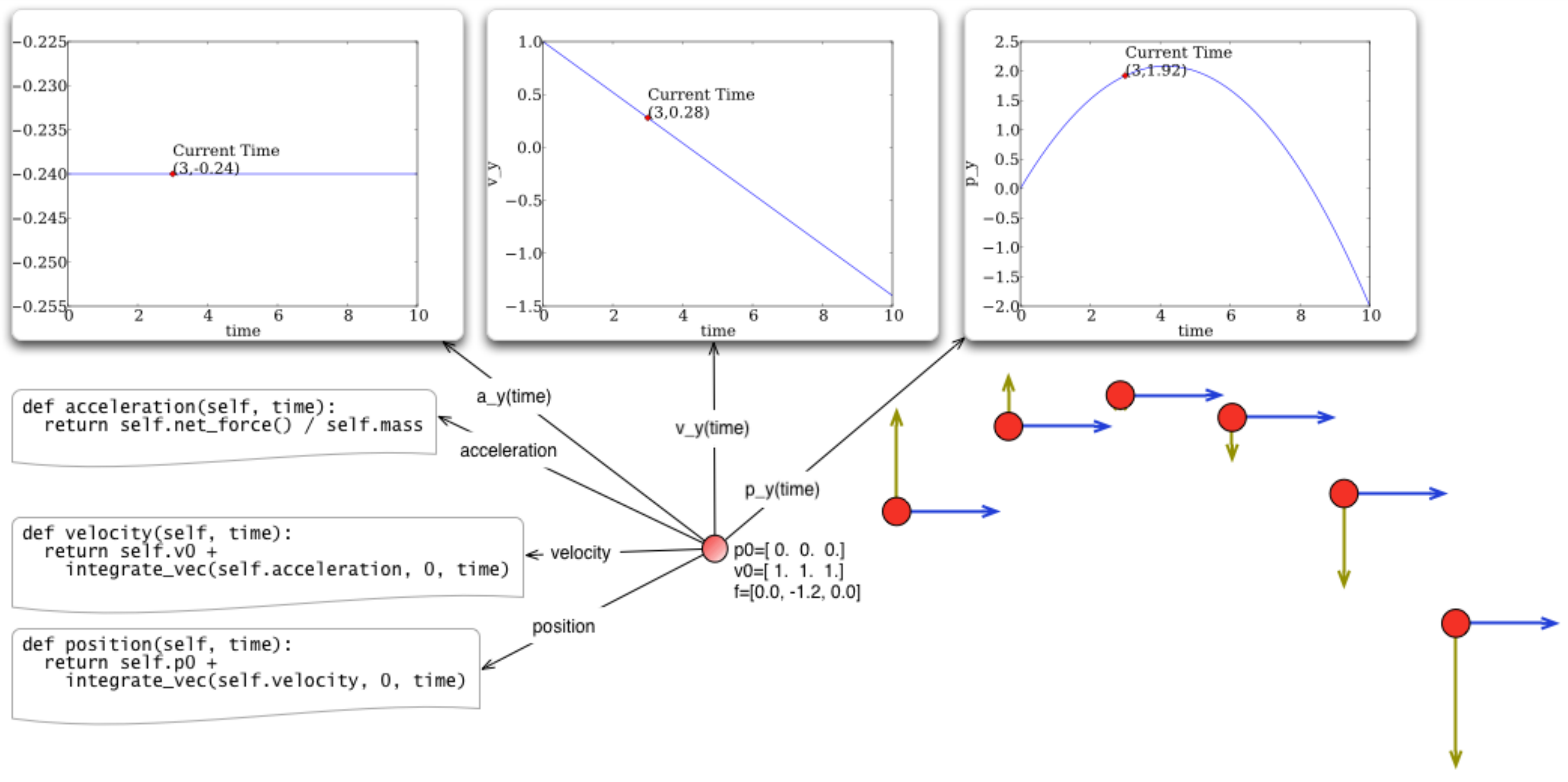}}
\caption{\texttt{Ball} in motion: functions of time as code, graphs, animation \DUrole{label}{phyfig}}
\end{figure*}

Below is a model of the motion of a ball under constant force. The ball has vector-valued attributes for initial position, velocity, and forces (lines 2,3). The functions \texttt{acceleration}, \texttt{velocity}, and \texttt{position} are pure functions of time and use numerical integration. We visualize ball \texttt{b} via \texttt{showGraph} and \texttt{animate} (lines 18-19). Like all visualizations, the animation is specified by a \emph{template} (line 21): a dictionary of visual properties, except that these properties can be \emph{functions} of the \emph{object} being animated, and the \emph{time} at which its attributes values are computed.\begin{Verbatim}[commandchars=\\\{\},numbers=left,firstnumber=1,stepnumber=1,fontsize=\footnotesize,xleftmargin=2.25mm,numbersep=3pt]
\PY{k}{class} \PY{n+nc}{Ball}\PY{p}{(}\PY{n}{Concept}\PY{p}{)}\PY{p}{:}
  \PY{n}{mass}\PY{p}{,} \PY{n}{p0}\PY{p}{,} \PY{n}{v0} \PY{o}{=} \PY{n}{Float}\PY{p}{,} \PY{n}{Instance}\PY{p}{(}\PY{n}{vector}\PY{p}{)}\PY{p}{,} \PY{o}{.}\PY{o}{.}\PY{o}{.}
  \PY{n}{forces} \PY{o}{=} \PY{n}{List}\PY{p}{(}\PY{n}{vector}\PY{p}{)}
  \PY{k}{def} \PY{n+nf}{net\PYZus{}force}\PY{p}{(}\PY{n+nb+bp}{self}\PY{p}{)}\PY{p}{:}
    \PY{k}{return} \PY{n}{v\PYZus{}sum}\PY{p}{(}\PY{n+nb+bp}{self}\PY{o}{.}\PY{n}{forces}\PY{p}{)}
  \PY{k}{def} \PY{n+nf}{acceleration}\PY{p}{(}\PY{n+nb+bp}{self}\PY{p}{,} \PY{n}{time}\PY{p}{)}\PY{p}{:}
    \PY{k}{return} \PY{n+nb+bp}{self}\PY{o}{.}\PY{n}{net\PYZus{}force}\PY{p}{(}\PY{p}{)} \PY{o}{/} \PY{n+nb+bp}{self}\PY{o}{.}\PY{n}{mass}
  \PY{k}{def} \PY{n+nf}{velocity}\PY{p}{(}\PY{n+nb+bp}{self}\PY{p}{,} \PY{n}{time}\PY{p}{)}\PY{p}{:}
    \PY{k}{return} \PY{n+nb+bp}{self}\PY{o}{.}\PY{n}{v0} \PY{o}{+} \PY{n}{v\PYZus{}integrate}\PY{p}{(}\PY{n+nb+bp}{self}\PY{o}{.}\PY{n}{acceleration}\PY{p}{,} \PY{n}{time}\PY{p}{)}
  \PY{k}{def} \PY{n+nf}{position}\PY{p}{(}\PY{n+nb+bp}{self}\PY{p}{,} \PY{n}{time}\PY{p}{)}\PY{p}{:}
    \PY{k}{return} \PY{n+nb+bp}{self}\PY{o}{.}\PY{n}{p0} \PY{o}{+} \PY{n}{v\PYZus{}integrate}\PY{p}{(}\PY{n+nb+bp}{self}\PY{o}{.}\PY{n}{velocity}\PY{p}{,} \PY{n}{time}\PY{p}{)}

  \PY{k}{def} \PY{n+nf}{p\PYZus{}x}\PY{p}{(}\PY{n+nb+bp}{self}\PY{p}{,} \PY{n}{time}\PY{p}{)}\PY{p}{:} \PY{o}{.}\PY{o}{.}\PY{o}{.}\PY{o}{.}
  \PY{k}{def} \PY{n+nf}{p\PYZus{}y}\PY{p}{(}\PY{n+nb+bp}{self}\PY{p}{,} \PY{n}{time}\PY{p}{)}\PY{p}{:} \PY{o}{.}\PY{o}{.}\PY{o}{.}\PY{o}{.}

\PY{n}{b} \PY{o}{=} \PY{n}{Ball}\PY{p}{(}\PY{n}{p0}\PY{o}{=}\PY{o}{.}\PY{o}{.}\PY{o}{.}\PY{p}{,} \PY{n}{v0}\PY{o}{=}\PY{o}{.}\PY{o}{.}\PY{o}{.}\PY{p}{,} \PY{n}{mass}\PY{o}{=}\PY{o}{.}\PY{o}{.}\PY{o}{.}\PY{p}{,} \PY{n}{forces}\PY{o}{=}\PY{o}{.}\PY{o}{.}\PY{o}{.}\PY{p}{)}
\PY{n}{m} \PY{o}{=} \PY{n}{Model}\PY{p}{(}\PY{n}{b}\PY{p}{)}
\PY{n}{m}\PY{o}{.}\PY{n}{showGraph}\PY{p}{(}\PY{n}{b}\PY{p}{,} \PY{p}{(}\PY{l+s}{\PYZsq{}}\PY{l+s}{a\PYZus{}y}\PY{l+s}{\PYZsq{}}\PY{p}{,}\PY{l+s}{\PYZsq{}}\PY{l+s}{v\PYZus{}y}\PY{l+s}{\PYZsq{}}\PY{p}{,}\PY{l+s}{\PYZsq{}}\PY{l+s}{p\PYZus{}y}\PY{l+s}{\PYZsq{}}\PY{p}{)}\PY{p}{,} \PY{p}{(}\PY{l+m+mi}{0}\PY{p}{,}\PY{l+m+mi}{10}\PY{p}{)}\PY{p}{)}
\PY{n}{m}\PY{o}{.}\PY{n}{animate}\PY{p}{(}\PY{n}{b}\PY{p}{,}
    \PY{p}{(}\PY{l+m+mi}{0}\PY{p}{,}\PY{l+m+mi}{10}\PY{p}{)}\PY{p}{,}
    \PY{p}{[}\PY{p}{\PYZob{}}\PY{n}{K}\PY{o}{.}\PY{n}{new}\PY{p}{:} \PY{n}{K}\PY{o}{.}\PY{n}{shape}\PY{p}{,}
      \PY{n}{K}\PY{o}{.}\PY{n}{origin}\PY{p}{:} \PY{k}{lambda} \PY{n}{b}\PY{p}{,}\PY{n}{t}\PY{p}{:} \PY{p}{[}\PY{n}{b}\PY{o}{.}\PY{n}{p\PYZus{}x}\PY{p}{(}\PY{n}{t}\PY{p}{)}\PY{p}{,} \PY{n}{b}\PY{o}{.}\PY{n}{p\PYZus{}y}\PY{p}{(}\PY{n}{t}\PY{p}{)}\PY{p}{]}\PY{p}{]}\PY{p}{\PYZcb{}}\PY{p}{,}
     \PY{p}{\PYZob{}}\PY{n}{K}\PY{o}{.}\PY{n}{new}\PY{p}{:} \PY{n}{K}\PY{o}{.}\PY{n}{line}\PY{p}{,} \PY{n}{point\PYZus{}list}\PY{o}{=}\PY{k}{lambda} \PY{n}{b}\PY{p}{,}\PY{n}{t}\PY{p}{:} \PY{o}{.}\PY{o}{.}\PY{o}{.}\PY{p}{\PYZcb{}}\PY{p}{,}
     \PY{p}{\PYZob{}}\PY{n}{K}\PY{o}{.}\PY{n}{new}\PY{p}{:} \PY{n}{K}\PY{o}{.}\PY{n}{line}\PY{p}{,} \PY{n}{point\PYZus{}list}\PY{o}{=}\PY{k}{lambda} \PY{n}{b}\PY{p}{,}\PY{n}{t}\PY{p}{:} \PY{o}{.}\PY{o}{.}\PY{o}{.}\PY{p}{\PYZcb{}}\PY{p}{]} \PY{p}{)}
\end{Verbatim}
PySTEMM generates graphs of the time-varying functions, and a 2-D animation of the position and velocity vectors of the ball over time (Figure \DUrole{ref}{phyfig}).

\section{Engineering%
  \label{engineering}%
}
\begin{figure}[]\noindent\makebox[\columnwidth][c]{\includegraphics[scale=0.50]{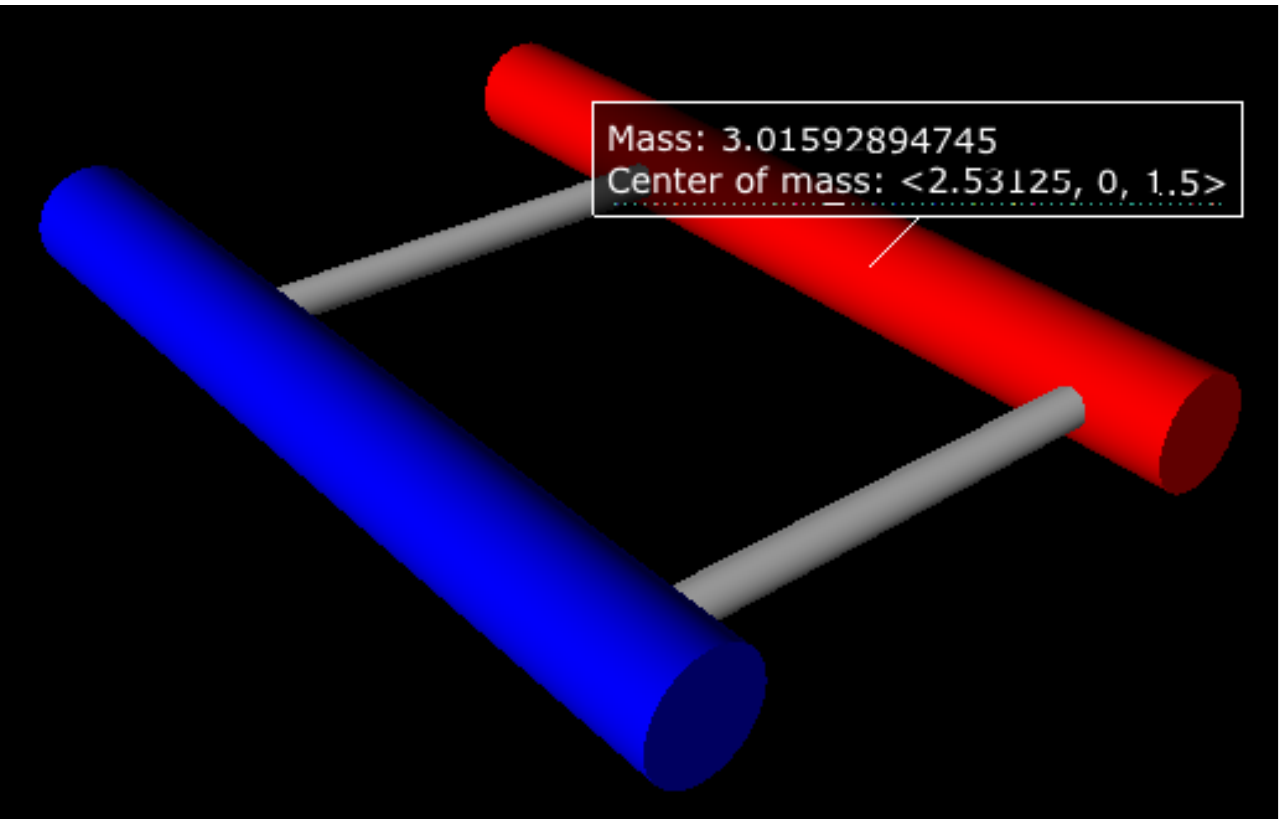}}
\caption{\texttt{ROV} made of \texttt{PVCPipes}. \DUrole{label}{rovfig}}
\end{figure}

In Summer 2012 I attended the OEX program at MIT, where we designed and built a marine remote-operated vehicle (ROV) with sensors to monitor water conditions. I later used PySTEMM to recreate models of the ROV, and generate engineering attributes and 3-D visualizations like Figure \DUrole{ref}{rovfig}.

The \texttt{ROV} is built from \texttt{PVCPipes} in a functional style. To create several \texttt{PVCPipes} positioned and sized relative to each other, the model uses pure functions like \texttt{shift} and \texttt{rotate} that take a \texttt{PVCPipe} and some geometry, and produce a transformed \texttt{PVCPipe}. This makes it simple to define parametric models and rapidly try different \texttt{ROV} structures. The model shown excludes motors, micro-controller, and computed drag, net force, and torque.\begin{Verbatim}[commandchars=\\\{\},fontsize=\footnotesize]
\PY{k}{class} \PY{n+nc}{PVCPipe}\PY{p}{(}\PY{n}{Concept}\PY{p}{)}\PY{p}{:}
  \PY{n}{length}\PY{p}{,} \PY{n}{radius}\PY{p}{,} \PY{n}{density} \PY{o}{=} \PY{n}{Float}\PY{p}{,} \PY{n}{Float}\PY{p}{,} \PY{n}{Float}
  \PY{k}{def} \PY{n+nf}{shift}\PY{p}{(}\PY{n+nb+bp}{self}\PY{p}{,} \PY{n}{v}\PY{p}{)}\PY{p}{:}
    \PY{k}{return} \PY{n}{PVCPipe}\PY{p}{(}\PY{n+nb+bp}{self}\PY{o}{.}\PY{n}{p0} \PY{o}{+} \PY{n}{v}\PY{p}{,} \PY{n+nb+bp}{self}\PY{o}{.}\PY{n}{r}\PY{p}{,} \PY{n+nb+bp}{self}\PY{o}{.}\PY{n}{axis}\PY{p}{)}
  \PY{k}{def} \PY{n+nf}{rotate}\PY{p}{(}\PY{n+nb+bp}{self}\PY{p}{,} \PY{n}{a}\PY{p}{)}\PY{p}{:}
    \PY{k}{return} \PY{n}{PVCPipe}\PY{p}{(}\PY{n+nb+bp}{self}\PY{o}{.}\PY{n}{p0}\PY{p}{,} \PY{n+nb+bp}{self}\PY{o}{.}\PY{n}{r}\PY{p}{,} \PY{n+nb+bp}{self}\PY{o}{.}\PY{n}{axis} \PY{o}{+} \PY{n}{a}\PY{p}{)}

\PY{k}{class} \PY{n+nc}{ROV}\PY{p}{(}\PY{n}{Concept}\PY{p}{)}\PY{p}{:}
  \PY{n}{body} \PY{o}{=} \PY{n}{List}\PY{p}{(}\PY{n}{PVCPipe}\PY{p}{)}
  \PY{k}{def} \PY{n+nf}{mass}\PY{p}{(}\PY{n+nb+bp}{self}\PY{p}{)}\PY{p}{:} \PY{o}{.}\PY{o}{.}\PY{o}{.}
  \PY{k}{def} \PY{n+nf}{center\PYZus{}of\PYZus{}mass}\PY{p}{(}\PY{n+nb+bp}{self}\PY{p}{)}\PY{p}{:} \PY{o}{.}\PY{o}{.}\PY{o}{.}
  \PY{k}{def} \PY{n+nf}{moment\PYZus{}of\PYZus{}inertia}\PY{p}{(}\PY{n+nb+bp}{self}\PY{p}{)}\PY{p}{:} \PY{o}{.}\PY{o}{.}\PY{o}{.}

\PY{n}{p1} \PY{o}{=} \PY{n}{PVCPipe}\PY{p}{(}\PY{o}{.}\PY{o}{.}\PY{o}{.}\PY{o}{.}\PY{p}{)}
\PY{n}{p2} \PY{o}{=} \PY{n}{p1}\PY{o}{.}\PY{n}{shift}\PY{p}{(}\PY{p}{(}\PY{l+m+mi}{0}\PY{p}{,}\PY{l+m+mi}{0}\PY{p}{,}\PY{l+m+mi}{3}\PY{p}{)}\PY{p}{,} \PY{o}{.}\PY{o}{.}\PY{o}{.}\PY{p}{)}
\PY{n}{c1}\PY{p}{,} \PY{n}{c2} \PY{o}{=} \PY{n}{p1}\PY{o}{.}\PY{n}{rotate}\PY{p}{(}\PY{p}{(}\PY{l+m+mi}{0}\PY{p}{,}\PY{l+m+mi}{0}\PY{p}{,}\PY{l+m+mi}{90}\PY{p}{)}\PY{p}{)}\PY{o}{.}\PY{o}{.}\PY{o}{.}
\PY{n}{rov} \PY{o}{=} \PY{n}{ROV}\PY{p}{(}\PY{n}{body}\PY{o}{=}\PY{n}{p1}\PY{p}{,} \PY{n}{p2}\PY{p}{,} \PY{n}{c1}\PY{p}{,} \PY{n}{c2}\PY{p}{)}
\end{Verbatim}




\section{Implementation%
  \label{implementation}%
}

\subsection{Architecture%
  \label{architecture}%
}
The overall architecture of PySTEMM, illustrated in Figure \DUrole{ref}{archfig}, has two main parts: \emph{Tool} and \emph{Model Library}. The \emph{tool} manipulates \emph{models}, traversing them at the type and instance level and generating visualizations. The \emph{model library} includes the models presented in this paper and any additional models any PySTEMM user would create. The \emph{tool} is implemented with 3 classes:%
\begin{itemize}

\item 

\texttt{Concept}: a superclass that triggers special handling of the concept type to process attribute-type definitions.
\item 

\texttt{Model}: a collection of concepts classes and concept instances, configured with some visualization.
\item 

\texttt{View}: an interface to a drawing application scripted via AppleScript.
\end{itemize}

Figure \DUrole{ref}{archfig} explains the architecture in more detail, and lists external modules that were used for specific purposes. PySTEMM uses the Enthought \texttt{traits} module \cite{Tra14} to define attribute types for a concept. Traits provides a class \texttt{HasTraits} with a custom meta-class, and pre-defined traits such as \texttt{List}, \texttt{Tuple}, \texttt{String}, and \texttt{Int}. The \texttt{Concept} class derives from \texttt{HasTraits}, which triggers \texttt{traits} to capture concept attribute type definitions and generate constructor and attribute logic for checked attribute assignment.
\begin{figure*}[]\noindent\makebox[\textwidth][c]{\includegraphics[scale=0.40]{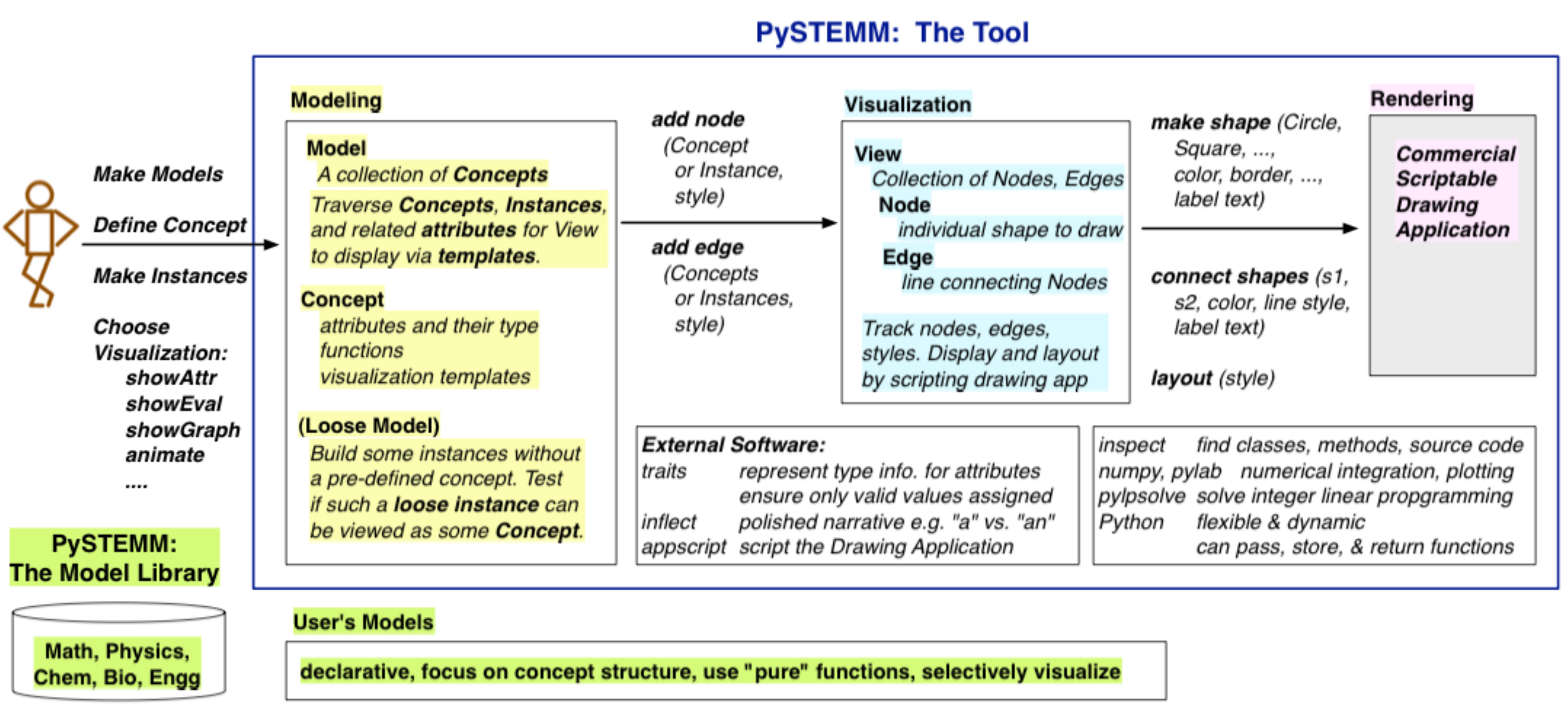}}
\caption{Architecture of PySTEMM. \DUrole{label}{archfig}}
\end{figure*}

We gain several benefits by building models with immutable objects and pure functions:%
\begin{itemize}

\item 

The \emph{user models} can be manipulated by the \emph{tool} more easily to provide tool capabilities like animation and graph-plotting, based on evaluating pure functions at different points in time.
\item 

The values of computed attributes and other intermediate values can be visualized as easily and unambiguously as any stored attributes.
\item 

Debugging becomes much less of an issue since values do not change while executing a model, and the definitions parallel the math taught in school science.
\end{itemize}

The source code for PySTEMM is available at \url{https://github.com/kdz/pystemm}.

\subsection{Python%
  \label{python}%
}

Python provides many advantages to this project:%
\begin{itemize}

\item 

adequate support for high-order functions and functional programming;
\item 

lightweight and flexible syntax, with convenient modeling constructs like lists, tuples, and dictionaries;
\item 

good facilities to manipulate classes, methods, and source code;
\item 

vast ecosystem of open-source libraries, including excellent ones for scientific computing.
\end{itemize}

\subsection{Templates%
  \label{templates}%
}

All visualization is defined by \emph{templates} containing visual property values, or functions to compute those values:\begin{Verbatim}[commandchars=\\\{\},fontsize=\footnotesize]
\PY{n}{Concept\PYZus{}Template} \PY{o}{=} \PY{p}{\PYZob{}}
  \PY{n}{K}\PY{o}{.}\PY{n}{text}\PY{p}{:} \PY{k}{lambda} \PY{n}{concept}\PY{p}{:} \PY{n}{computeClassLabel}\PY{p}{(}\PY{n}{concept}\PY{p}{)}\PY{p}{,}
  \PY{n}{K}\PY{o}{.}\PY{n}{name}\PY{p}{:} \PY{l+s}{\PYZsq{}}\PY{l+s}{Rectangle}\PY{l+s}{\PYZsq{}}\PY{p}{,}
  \PY{n}{K}\PY{o}{.}\PY{n}{corner\PYZus{}radius}\PY{p}{:} \PY{l+m+mi}{6}\PY{p}{,}
  \PY{o}{.}\PY{o}{.}\PY{o}{.}
  \PY{n}{K}\PY{o}{.}\PY{n}{gradient\PYZus{}color}\PY{p}{:} \PY{l+s}{\PYZdq{}}\PY{l+s}{Snow}\PY{l+s}{\PYZdq{}}\PY{p}{\PYZcb{}}
\end{Verbatim}
The primary operation on a template is to \emph{apply} it to some modeling object, typically a concept class or instance:\begin{Verbatim}[commandchars=\\\{\},fontsize=\footnotesize]
\PY{k}{def} \PY{n+nf}{apply\PYZus{}template}\PY{p}{(}\PY{n}{t}\PY{p}{,} \PY{n}{obj}\PY{p}{,} \PY{n}{time}\PY{o}{=}\PY{n+nb+bp}{None}\PY{p}{)}\PY{p}{:}
  \PY{c}{\PYZsh{} t.values are drawing\PYZhy{}app values, or functions}
  \PY{c}{\PYZsh{} obj: any object, passed into template functions}
  \PY{c}{\PYZsh{} returns copy of t, F(obj) replaces functions F}
  \PY{k}{if} \PY{n+nb}{isinstance}\PY{p}{(}\PY{n}{t}\PY{p}{,} \PY{n+nb}{dict}\PY{p}{)}\PY{p}{:}
    \PY{k}{return} \PY{p}{\PYZob{}}\PY{n}{k}\PY{p}{:} \PY{n}{apply\PYZus{}template}\PY{p}{(}\PY{n}{v}\PY{p}{,} \PY{n}{obj}\PY{p}{,} \PY{n}{time}\PY{p}{)}
               \PY{k}{for} \PY{n}{k}\PY{p}{,} \PY{n}{v} \PY{o+ow}{in} \PY{n}{t}\PY{o}{.}\PY{n}{items}\PY{p}{(}\PY{p}{)}\PY{p}{\PYZcb{}}
  \PY{k}{if} \PY{n+nb}{isinstance}\PY{p}{(}\PY{n}{t}\PY{p}{,} \PY{n+nb}{list}\PY{p}{)}\PY{p}{:}
    \PY{k}{return} \PY{p}{[}\PY{n}{apply\PYZus{}template}\PY{p}{(}\PY{n}{x}\PY{p}{,} \PY{n}{obj}\PY{p}{,} \PY{n}{time}\PY{p}{)}
               \PY{k}{for} \PY{n}{x} \PY{o+ow}{in} \PY{n}{t}\PY{p}{]}
  \PY{k}{if} \PY{n+nb}{callable}\PY{p}{(}\PY{n}{t}\PY{p}{)}\PY{p}{:}
    \PY{k}{return} \PY{n}{t}\PY{p}{(}\PY{n}{obj}\PY{p}{)} \PY{k}{if} \PY{n}{arity}\PY{p}{(}\PY{n}{t}\PY{p}{)}\PY{o}{==}\PY{l+m+mi}{1} \PY{k}{else} \PY{n}{t}\PY{p}{(}\PY{n}{obj}\PY{p}{,} \PY{n}{time}\PY{p}{)}
  \PY{k}{return} \PY{n}{t}
\end{Verbatim}
Animation templates have special case handling, since their functions take two parameters: the \emph{instance} to be rendered, and the \emph{time} at which to render its attributes.

Templates can also be \emph{merged}. Figure \DUrole{ref}{funcinstances} shows an  instance of \texttt{TableFunction} as a circle in the same color as the \texttt{TableFunction} class, by merging an \texttt{instance\_template} with a \texttt{class\_template}.

\section{Summary%
  \label{summary}%
}

I have described PySTEMM as a tool, model library, and approach for building executable concept models for a variety of STEM subjects. The PySTEMM approach, using immutable objects and pure functions in Python, can apply to all STEM areas. It supports learning through pictures, narrative, animation, and graph plots, all generated from a single model definition, with minimal incidental complexity and code debugging issues. Such modeling, if given a more central role in K-12 STEM education, could make STEM learning much more deeply engaging.





\end{document}